\documentclass[lineno]{jfm}

\usepackage{graphicx}
\usepackage{newtxtext}
\usepackage{newtxmath}
\usepackage{natbib}
\usepackage{hyperref}
\usepackage{cancel}
\usepackage{rotating} 

\usepackage{overpic}
\usepackage{pict2e} 

\usepackage{svg} 
\usepackage{relsize} 
\usepackage{xcolor}
\usepackage{todonotes}

\definecolor{deeppurple}{RGB}{75, 0, 130}

\hypersetup{
    colorlinks = true,
    urlcolor   = blue,
    citecolor  = black,
}

\newcommand{\RomanNumeralCaps}[1]
\linenumbers

\newcommand{\pd}[2]{\frac{\partial #1}{\partial #2}}
\newcommand{\f}[2]{\frac{#1}{#2}} 
\newcommand\Sty{\mbox{$\mathrm{St}$}} 

\newcommand\abeqn[2]{\refstepcounter{equation}
\[
\label{#1}
#2
\eqno{(\theequation\textit{a,b})}
\]} 

\newcommand{\dd}[2]{\frac{{\mathrm d}#1}{{\mathrm d}#2}}

\newif\ifdraft
\draftfalse     

\usepackage[normalem]{ulem} 
\usepackage{xcolor}

\newcommand{\change}[2]{%
  \ifdraft
    \textcolor{red}{\sout{#1}}%
    \textcolor{blue}{#2}%
  \else
    #2%
  \fi
}

\newcommand{\delete}[1]{%
  \ifdraft
    \textcolor{red}{\sout{#1}}%
  \else
  \fi
}

\newcommand{\add}[1]{%
  \ifdraft
    \textcolor{blue}{#1}%
  \else
    #1%
  \fi
}

\newcommand{\ownadd}[1]{%
  \ifdraft
    \textcolor{teal}{#1}%
  \else
    #1%
  \fi
}

\title{A multiple-scales framework for branched channel filters}

\author{T. Fastnedge \corresp{\email{torin.fastnedge@maths.ox.ac.uk}},
 C. J. W. Breward 
 \and I. M. Griffiths}

\affiliation{Mathematical Institute, University of Oxford, Woodstock Road, Oxford, OX2 6GG, UK}

\begin{document}
\maketitle

\noindent Fibres shed from our clothes during a washing machine cycle constitute around $35\%$ of the primary microplastics in our oceans. Current conventional dead-end washing machine filters clog relatively quickly and require frequent cleaning. We consider a new concept, \emph{ricochet separation}, inspired by the feeding process of manta rays, to reduce the cleaning frequency. In such a device, some fluid is diverted through branched channels whilst particles ricochet off the wall structure, forcing them back into the main flow and then into the dead-end filter.

In this paper, we \add{use this industrially inspired challenge to motivate the study of} \delete{consider} a simple branched\change{-}{ }channel filter \delete{structure} beneath a high-Reynolds-number laminar flow, in the case where the branch separation is much larger than the thickness of the viscous boundary layer. We use multiple-scales techniques to derive an effective leakage boundary condition, which smooths out localised effects in the flow velocity and pressure that arise due to the discrete branched channels, and then use this boundary condition to explicitly determine the flow away from the boundary. We find that our explicit solution compares well with an analogous numerical solution containing a discrete set of branched channels.

We further consider the behaviour of individual spherical particles in the device, with their trajectories determined via a simple force balance model with a wall-bounce condition. We explore the dependence of the fraction of particles that flow into the branched channels on the \delete{particle's} Stokes number. The resulting combined model is able to predict the relationship between the efficiency of a ricochet filter device and the design and operating parameters, avoiding the need to conduct extensive numerically challenging simulations.

\bigskip

\begin{keywords}
Authors should not enter keywords on the manuscript.
\end{keywords}

\section{Introduction}
\label{section:Introduction}

In a study by \cite{InternationalUnionReport} for the International Union for Conservation of Nature, it was found that between $15 \%$ and $31 \%$ of all plastics in the ocean originate from household and industrial appliances, termed \emph{primary microplastics}, and up to $35 \%$ of these \add{particles} are composed of microfibres from our clothes. The term `microplastic' was introduced by \cite{thompson2004lostatsea}, who identified the danger of \change{these}{the build up of these} micron-sized plastic\delete{s building up} \add{particles} in the oceans, causing a risk to marine life. \cite{napper2016release} and \cite{deFalco2019contribution} found that the microplastics removed from textiles take the form of microfibres, with diameters of $11.9$--$17.7 $$\mu$m
and lengths of $360$--$660$$\mu$m. Recent work on quantifying the number of microfibres that end up in the ocean after each wash has been reviewed by several authors, with a consensus of at least $700,000$ \citep{dris2015beyond, napper2016release, acharya2021microfibers, QuantificationOfMicrofibres}.

\begin{figure}
    \centering
    \vspace{3mm}
    \begin{overpic}[width=1\textwidth,tics=10]{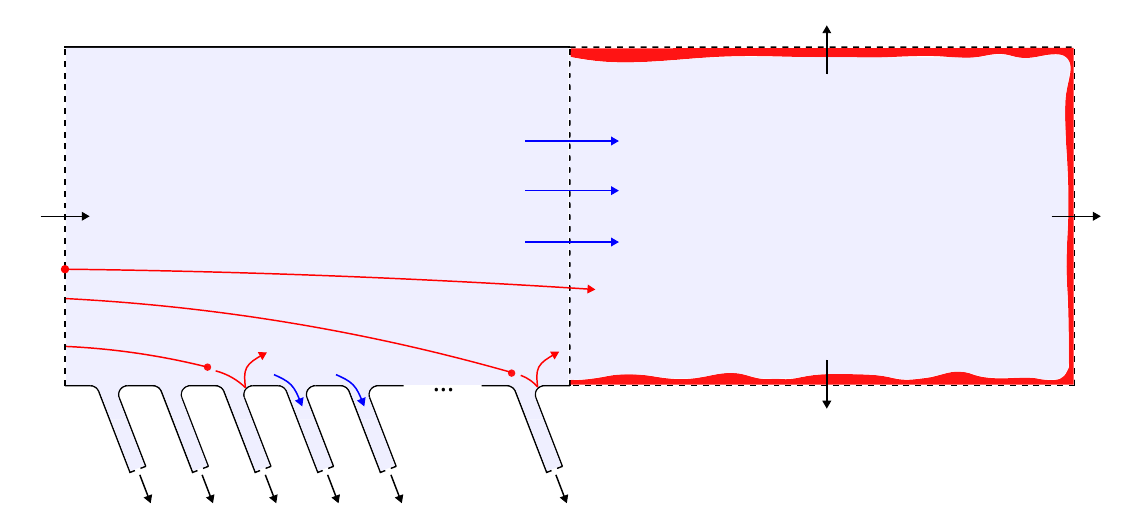}
    \put(67,28){Dead-end filter}
    \put(21,29.5){Ricochet device}
    \put(17,26.5){(Branched channel filter)}
    \end{overpic} 
    \vspace{1mm}
    \caption{Layout schematic of a branched channel filter preceding a dead-end filter. Microfibre particles,  trajectories and foulant are indicated in red and water flow is indicated in blue. The operating directions are indicated by black arrows.}
    \label{fig:Ricochet}
\end{figure}

It is becoming vital to remove these fibres at the source of the discharge, \textit{i.e.,} from washing machine wastewater. Some conventional washing machines incorporate dead-end filters which are effective at removing microfibres but clog relatively quickly \citep{enten2020optimizing, akarsu2021removal}. \cite{enten2020optimizing} also examine periodic back-flushing through the dead-end filter to wash fibres off the surface and reduce the need for filter cleaning. However, consumers often postpone cleaning such filters for as long as possible, which results in the fluid bypassing the filter through the emergency overflow mechanism, thus negating the purpose of the filter. An alternative method for increasing the lifespan of a washing machine dead-end filter, being explored by Beko PLC, involves diverting as much water away from the dead-end filter as possible whilst retaining as many microfibre particles as possible flowing into the dead-end filter (see figure~\ref{fig:Ricochet}).

Beko's prototype designs are motivated by a fluid--particle separation technique identified in manta ray fish, a type of suspension feeder, coining the term \emph{ricochet separation} \citep{MantaRayRicochetSeparation}. Within the manta ray's mouth, there is a gill-like pore structure resembling branched channels. Plankton-rich water flows over this structure, with clean water flowing through the pores while plankton collide\delete{s} with the rigid structure and ricochet\delete{s} back into the main free-stream flow above the pores, thus providing an efficient filtration mechanism. \citet{MantaRayRicochetSeparation} conclude that filtration efficiency increases for \change{large}{high}-Reynolds\add{-}number flows, and that particles much smaller than the size of the pores may be filtered, with increasing efficiency for increasing size. They simulate a quasi-steady flow and model the particle dynamics using Newton's second law of motion, and compare the results with experiments. 

Beko believe that the ricochet method might be an excellent way for removal of a substantial amount of water whilst retaining microplastic particles, removing the challenge of constructing a device with pores sufficiently small to remove the particles via size exclusion. Their aim is to use ricochet separation prior to the dead-end filtration, to reduce the amount of water --- but not microfibres --- that flows into a dead-end filter --- a trade-off between maximising the flow and minimising the number of particles through the branched \delete{pore}\ownadd{channel} structure.

The method of ricochet separation in manta rays, and various other mobula rays, has been further considered by \cite{LeakyChannels}. They consider a framework to describe the trade-off between fluid leakage rate and the critical particle size through the branched channels. For low\add{-}Reynolds\add{-}numbers, they derive an analytic solution for the leaking rate through each branched channel, whereas for larger Reynolds numbers, they rely on numerical simulations. In the latter regime, numerical simulations indicate that altering the branch angle while keeping the branch width constant does not substantially affect the leakage rate. In their model, as in a mobula ray's mouth, they assume the pressure at all outlets is atmospheric. This differs from a scenario in which a dead-end filter is placed at the main outlet, as this will experience a pressure build-up. They provide a relationship between the flow leakage rate and particle cut-off size in both the low\add{-} and high\add{-}Reynolds\add{-}number cases, concluding that the particle cut-off size is smaller for smaller leakage rates. In a different approach, \cite{hamann2023suspension} designed and analysed \delete{`mucus' and} `semi'-cross-flow filters, also motivated by suspension feeders, to explore whether a single filter is possible for microplastic capture without the need for a further dead-end filter \ownadd{--- later proposing such a filter in \cite{hamann2025self}}. 

In this paper, \add{motivated by the industrial idea of ricochet separation for filtration, we} \delete{will} build and solve a mathematical model to understand the trade-off between leakage rate and particle size that avoids the need to perform computationally challenging numerical simulations in the high-Reynolds-number laminar regime. We will exploit the fact that there are many branched channels in the device and so the proportion of fluid that passes through each one is low, which will give rise to multi-scale behaviour. \delete{We will use a multiple scales analysis to derive an effective boundary condition that encapsulates the key behaviour in the device. Such a strategy has been used in other contexts. For instance, when describing Faraday cages, Chapman \textit{et al.} (2015) and Hewett \& Hewitt (2016), derive an effective boundary condition for the electric potential in a discrete metal cage, smoothing out the effects caused by individual point-sources. In a similar approach, Bruna \textit{et al.} (2015) use homogenisation to derive an effective diffusive flux condition for the concentration through a thin-film composite membrane over many pore structures.} 

\delete{To facilitate a mathematical approach, we consider a simpler geometry than Divi \textit{et al.}(2018) and Mao \textit{et al.} (2024), removing the rounded lobe structure and replacing this with a series of T-junction branches equi-distance apart along a channel. A similar set-up was studied by Dalwadi \textit{et al.} (2020), however, this was for a single T-junction with normal flow to the branched channel and was not subsequently homogenised, nor studied for tangential flow over the branched channel. Nevertheless, we will consider some of the techniques described in  Dalwadi \textit{et al.} (2020), while following the homogenisation techniques by Chapman \textit{et al.} (2015), approximating the branched channels as point-sinks, and using complex variable theory to conformally map and solve for the velocity potential. This will allow us to find an effective boundary condition for the flow through the branched channels.}

\delete{A similar effective boundary condition has also been considered for tangential flow over a wall with periodic imperfections by Bottaro \& Naqvi (2020), using homogenisation techniques to smooth out the flow over the complex geometry, recovering well understood phenomena such as Navier slip in their leading-order results. This is extended in the work by Naqvi (2021), where they further consider homogenisation of the flow over and through porous micro-structures and elastic surfaces, for which we consider similar ideas in this paper. This concept of asymptotic homogenisation was formally introduced by Bensoussan \textit{et al.} (1978), finding that when utilising asymptotic expansions in a microscale cell structure, they were able to find effective equations to describe the macroscale behaviour. This drastically simplifies the numerical complexity of problems which is made simpler via homogenisation. This is similar to the approach we take in this paper, however we do not find effective equations, rather effective boundary conditions, and so our method is instead a multi-scale and boundary layer analysis (Van Dyke 1975; Hinch 1991).}

\add{To facilitate a mathematical approach, we consider a simpler geometry than \cite{MantaRayRicochetSeparation} and \cite{LeakyChannels}, removing their rounded lobe structure and replacing this with a series of T-junction branched channels equidistantly spaced along a channel. We exploit the fact that there are many branched channels in the device and so the proportion of fluid that passes through each one is low, which gives rise to multi-scale behaviour. We use a multiple-scales analysis to derive an effective boundary condition that encapsulates the key behaviour in the device.}

\add{\cite{dalwadi2020mathematical} study the normal, but not tangential, flow in a single T-junction and they reduce the problem to that of a point-sink. We use this idea to transform the branched channels into a series of point-sinks of appropriate strengths. The idea of using homogenisation theory to derive effective boundary conditions is elegantly presented in \citep{chapman2015mathematics}, and further developed in \citep{hewett2016homogenized}, where effective boundary conditions for the electrical potential in a discrete metal cage is derived, which smooths out the effects caused by individual point-sources. We use the ideas in these papers to solve our resulting cell problem using complex variable theory to conformally map and solve for the velocity potential, which we then match into the outer flow to give us our effective leakage boundary condition.}

\add{The concept of asymptotic homogenisation was formally introduced by \cite{bensoussan2011asymptotic}. When using asymptotic expansions in a microscale cell structure, they were able to find effective equations to describe the macroscale behaviour, which incorporated microscale behaviour into the material parameters. This approach drastically simplifies the numerical complexity of problems. We take a similar approach in this paper; however, rather than deriving effective equations, we find effective boundary conditions, and so our method is instead a multi-scale and boundary layer analysis \citep{van1975perturbation, Hinch1991}.}

\add{Such an approach has also been considered by \cite{bruna2015effective} to derive an effective diffusive flux condition for the concentration through a thin-film composite membrane. Similarly, \cite{bottaro2020effective} have used homogenisation techniques to derive an effective boundary condition for purely tangential flow past a wall with periodic imperfections, with no fluid removal. Their result recovers well-understood phenomena such as Navier slip in their leading-order results. This is extended in the work by \cite{NaqviThesis}, where they further consider homogenisation of the flow over and through porous micro-structures and elastic surfaces. However, our set-up is different, since we have pressure gradients down our branched channels.}

We begin in \S\ref{section:Model_Derivation} by deriving a high-Reynolds-number laminar flow model in a branched channel filter. We simplify this model in \S\ref{section:Simplified_solution_structure} to an inviscid flow everywhere between two parallel fixed walls, with point-sinks along the bottom wall; the sink strength is related to the flux found via \change{Poiseulle}{Poiseuille} flow in each branched channel. We solve for the outer flow far away from the wall in \S\ref{section:outer}\delete{,} \add{and} the inner flow close to the wall in \S\ref{section:inner}\change{,}{;} \add{in \S\ref{sec:outerpart2} we return to find the full explicit outer solution} and \add{further} find composite solutions and an effective boundary condition in \S\ref{section:CompSolution}. We compare these asymptotic results with the numerical simulations in \S\ref{section:FlowResult}. In \S\ref{section:particle_model} and \S\ref{section:particle_results}, we consider a particle model, exploring how collisions with the wall take place under the influence of our asymptotic prediction for the flow. We model the motion of individual particles via a force balance model with a bouncing/removal wall condition to mimic ricochet or branched\change{-}{ }channel capture\add{,} respectively. For the application of our flow model in particle filtration, we explore the trade-off between the fraction of particles and the fluid flux fraction that passes through the branched\change{-}{ }channels. Finally, in \S\ref{section:conc} we discuss our findings, draw conclusions and provide extensions.

\section{Flow model derivation}
\label{section:Model_Derivation}

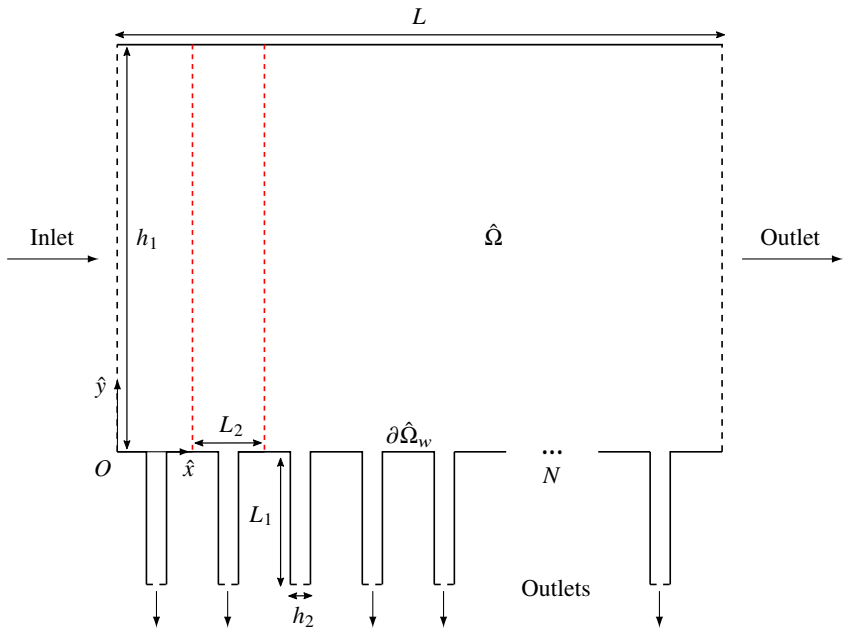
\begin{figure}
    \centering
    \vspace{3mm}
    \begin{overpic}[width=0.9\textwidth]{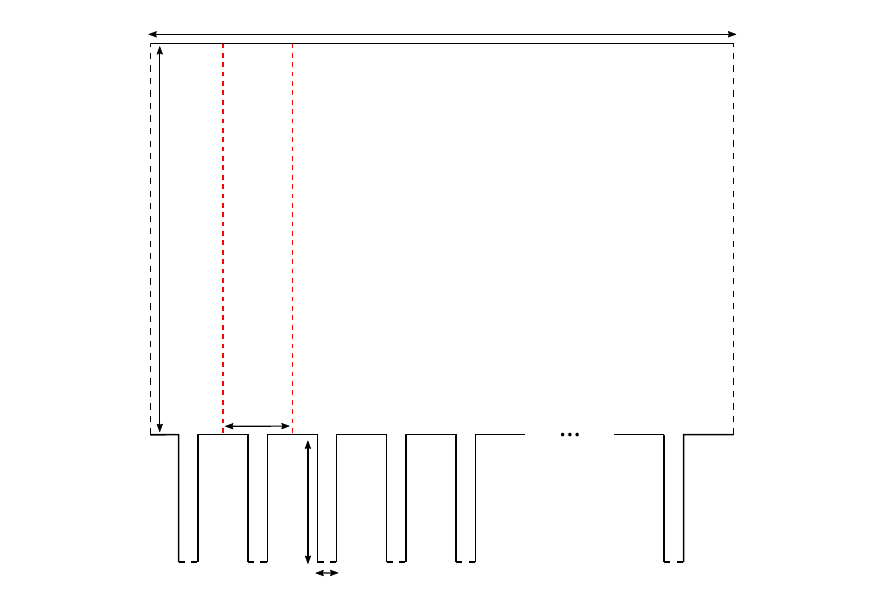}
    \put(36,-1){$h_2$}
    \put(28,19.75){$L_2$}
    \put(31.25,10){$L_1$}
    \put(19,40){$h_1$}
    \put(7.5,40){Inlet}
    \put(5,38.5){\vector(1,0){10}}
    \put(87,40){Outlet}
    \put(85,38.5){\vector(1,0){11}}
    \put(63.25,14.25){$N$}
    \put(61,2){Outlets}
    \put(21.25,2.5){\vector(0,-1){4}}
    \put(29,2.5){\vector(0,-1){4}}
    \put(44.8,2.5){\vector(0,-1){4}}
    \put(52.5,2.5){\vector(0,-1){4}}
    \put(76,2.5){\vector(0,-1){4}}
    \put(49,64){$L$}
    \put(57,40){$\hat{\Omega}$}
    \put(46.25,18.5){$\partial \hat{\Omega}_w$}
    %
    %
    \put(17,17.6){\vector(0,1){8}}
    \put(14.5,24){$\hat{y}$}
    \put(17,17.6){\vector(1,0){8}}
    \put(24.2,15){$\hat{x}$}
    \put(14.5,15){$O$}
    \end{overpic} 
    \vspace{1mm}
    \caption{$2$-dimensional repeatable T-junction domain, $\hat{\Omega}$, given by a main channel compartment with $N$ perpendicular branched channels on the bottom wall. Inlet and outlets are indicated by dashed black lines, the T-junction spacing is indicated by dashed red lines and boundary walls are denoted by $\partial \hat{\Omega}_w$, in solid black lines. The domain design parameters are indicated as $h_1$, $h_2$, $L$, $L_1$, $L_2$ and $N$.}
    \label{fig:Geometry}
\end{figure}

We consider a $2$-dimensional domain of length $L$ and inlet height $h_1$, with a series of $N$ repeated T-junctions of length $L_1$, width $h_2$, and spacing $L_2$, where $N \cdot L_2 = L$, protruding from the bottom wall (see figure \ref{fig:Geometry}). \add{Throughout this work, we use $\hat{\cdot}$\, to indicate dimensional quantities.} We use $\hat{x}$ to denote distance along the domain from the left-hand\delete{-} side, and $\hat{y}$ to denote distance above the bottom surface of the main channel. The centre of the top of each T-junction is located at $(\hat{x}_i, 0)$, where
\begin{equation}
    \hat{x}_i = \frac{(2 i - 1) L_2}{2},
\end{equation}
for $i = 1, 2, \cdots, N$, and we assume that the T-junctions are laterally symmetric about this centre point. We denote the full domain by $\hat{\Omega}$, as the union of the main channel $\left( \hat{x}, \hat{y} \right) \in [0,L] \times [0,h_1]$ and branched channels $\left( \hat{x}, \hat{y} \right) \in [\hat{x}_i-\frac{h_2}{2}, \hat{x}_i + \frac{h_2}{2}] \times [-L_1,0]$. The domain, $\hat{\Omega}$, has fixed boundary walls given by $\partial \hat{\Omega}_w$, along with \textit{inlet} at $\hat{x}=0$, \textit{main outlet} at $\hat{x}=1$, and \textit{outlets} at the bottom of the $N$ branched channels, $\hat{y}=-L_1$ (illustrated in figure~\ref{fig:Geometry}).

We consider a high-Reynolds-number, laminar, steady-state flow, in which the fluid velocity $\hat{\boldsymbol{u}} = (\hat{u}, \hat{v})$ and pressure $\hat{p}$ satisfy the steady\ownadd{, incompressible} Navier--Stokes equations,
\begin{align}
    \hat{\boldsymbol{\nabla}} \cdot \hat{\boldsymbol{u}} &= 0, \label{eqn:DimalConsMass}\\
   \rho_f (\hat{\boldsymbol{u}} \cdot \hat{\boldsymbol{\nabla}} ) \hat{\boldsymbol{u}} &= -  \hat{\boldsymbol{\nabla}} \hat{p} + \mu \change{\hat{\nabla^2}}{\hat{\nabla}^2} \hat{\boldsymbol{u}}, \label{eqn:DimalConsMom}
\end{align}
where $\rho_f$ is the fluid density, $\mu$ is  the viscosity. 
We impose a uniform plug inlet flow,
\begin{align}
    \hat{u} = \frac{\mathcal{Q}}{h_1}, \quad &\text{on} \quad \hat{x} = 0, \quad \hat{y} \in [0, h_1], \label{eqn:InletDimal}\\
    \hat{v} = 0, \quad &\text{on} \quad \hat{x} = 0, \quad \hat{y} \in [0, h_1], \label{eqn:Inletv0Dimal}
\end{align}
such that $\mathcal{Q}$ is the constant inlet flux at $\hat{x}=0$. We prescribe a no-slip and no-flux condition on the domain boundary, $\partial \hat{\Omega}_w$, given by
\begin{equation}
    \hat{\boldsymbol{u}} = \boldsymbol{0}.
    \label{eqn:DimalNoSlip}
\end{equation}
We apply an outlet pressure, given by
\begin{equation}
    \hat{p} = \hat{\mathcal{P}}_\text{out} + p_\text{atm}, \quad \text{on} \quad \hat{x} = L, \quad  \hat{y}\in [0, h_1], \label{eqn:FCOutletPressure}
\end{equation}
where $\hat{\mathcal{P}}_\text{out}$ is a given constant and $p_\text{atm}$ denotes constant atmospheric pressure. Finally, we impose that the pressure at the branch outlets is atmospheric and thus we write
\begin{equation}
    \hat{p} = p_\text{atm}, \quad \text{on} \quad \hat{y} = - L_1, \quad \hat{x} \in \bigcup_{i=1}^N \left[ \hat{x}_i - \frac{h_2}{2}, \hat{x}_i + \frac{h_2}{2} \right]. \label{eqn:AtmOutlet}
\end{equation}

\subsection{Non-dimensionalisation}

We non-dimensionalise the model, (\ref{eqn:DimalConsMass})--(\ref{eqn:AtmOutlet}), using the high-Reynolds-number scalings,
\begin{equation}
    \hat{\boldsymbol{x}} = L \boldsymbol{x}, \qquad \hat{\boldsymbol{u}} = \frac{\mathcal{Q}}{L} \boldsymbol{u}, \qquad \hat{p} = p_\text{atm} + \frac{\rho_f \mathcal{Q}^2}{\epsilon^{2} L^2} p, \label{eqn:FullScalings}
\end{equation}
where we have picked the pressure scaling so that the pressure at the end of the main outlet is $O(1)$, and
where
\begin{equation}
    \epsilon = \frac{L_2}{L}\delete{,}
\end{equation}
is defined as the dimensionless width of each T-junction, such that $N \cdot \epsilon = 1$. We \delete{will} assume that the hole-width-to-T-junction-width aspect ratio,
\begin{equation}
    \delta = \frac{h_2}{L_2} = \frac{h_2 N}{L},
\end{equation}
is a fixed constant. In this case, if we increase the number of branches, $N$, then the width of each T-junction decreases. Hence each individual branched channel of width $\delta \epsilon$\delete{,} will also decrease with $\epsilon$ and so a pressure increase of $1/\epsilon^2$ is required to force the same fluid flux through each hole. We define the Reynolds number, the dimensionless length of the branched channels, and the dimensionless height of the main channel as
\begin{equation}
    \Rey = \frac{\rho_f \mathcal{Q}}{\mu}, \qquad \lambda = \frac{L_1}{L}, \qquad \gamma = \frac{h_1}{L},
\end{equation}
respectively. We assume that in all cases, $\Rey \gg 1$, $\epsilon \ll 1$, $\gamma = \mathcal{O}(1)$, $\lambda = \mathcal{O}(1)$\add{,} and $\delta \epsilon \ll \lambda$.

The governing equations (\ref{eqn:DimalConsMass}) and (\ref{eqn:DimalConsMom}) become
\begin{align}
    \boldsymbol{\nabla} \cdot \boldsymbol{u} &= 0, \label{eqn:DimConsMass}\\
   (\boldsymbol{u} \cdot \boldsymbol{\nabla} ) \boldsymbol{u} &= - \frac{1}{\epsilon^2} \boldsymbol{\nabla} p + \frac{1}{\Rey} \nabla^2 \boldsymbol{u} \label{eqn:DimNS},
\end{align}
and the boundary conditions (\ref{eqn:InletDimal})--(\ref{eqn:AtmOutlet}) become
\begin{align}
    u = \frac{1}{\gamma}, \quad &\text{on} \quad x = 0, \quad y \in [0, \gamma],\label{eqn:dimlessInlet}\\
    v = 0, \quad &\text{on} \quad x = 0, \quad y \in [0, \gamma],\label{eqn:dimlessInletv0}\\
     u = v = 0, \quad &\text{on} \quad \partial \Omega_w,
    \label{eqn:DimlessSlip}\\
    p = \mathcal{P}_\text{out}, \quad &\text{on} \quad {x} = 1, \quad y \in [0, \gamma], \label{eqn:dimlessPoutCond}\\
    p = 0, \quad &\text{on} \quad y = - \lambda, \quad x \in \bigcup_{i=1}^N \left[ x_i - \frac{\delta \epsilon}{2}, x_i + \frac{\delta \epsilon}{2} \right], \label{eqn:AtmospDimless}
\end{align}
where
\begin{equation}
    x_i = \frac{(2 i - 1) \epsilon}{2},
    \label{eqn:DimlessCentreBranch}
\end{equation}
for $i = 1, 2, \cdots, N$, and
\begin{equation}
    \mathcal{P}_\text{out} = \frac{\epsilon^2 L^2}{\rho_f \mathcal{Q}^2} \hat{\mathcal{P}}_\text{out},
\end{equation}
which we assume is $\mathcal{O}(1)$.

\section{Flow solution structure}

\label{section:Simplified_solution_structure}

We begin the solution procedure by finding the flux through each branched channel via the well-known problem of unidirectional fully developed flow.

\subsection{Branched channel flow}

\label{section:BranchedFlow}

The dimensionless Navier--Stokes equations given by equations~(\ref{eqn:DimConsMass}) and (\ref{eqn:DimNS}) hold in each individual branched channel. Since we assume that $\delta \epsilon \ll \lambda$, each branched channel is long and thin, driven by a constant pressure drop with no-slip on the walls due to viscous effects. The inlet, outlet and no-slip boundary conditions on an isolated branched channel are given by
\begin{align}
    p = p(x_i, 0), \quad &\text{on} \quad y = 0, \label{eqn:InletChannelPressure}\\
    p = 0, \quad &\text{on} \quad y = - \lambda,\\
    \boldsymbol{u} = \boldsymbol{0}, \quad &\text{on} \quad x = x_i \pm \frac{\delta \epsilon}{2}, \label{eqn:NoSlipChannel}
\end{align}
respectively, where we assume that $p(x_i, 0)$ is the constant pressure at the top and centre of each branched channel via Taylor expansion. We note that there will be a small region (see Appendix \ref{appA}) near the top of the T-junction where the flow develops into unidirectional flow, but we neglect this here. The leading-order solution for channel flow with no-slip on the walls is
\begin{align}
    p &= p(x_i, 0) \left( 1 + \frac{y}{\lambda} \right),\\
    u &= 0,\\
    v &= \frac{\Rey}{2 \epsilon^2} \frac{\mathrm{d} p}{\mathrm{d} y} \left( (x - x_i)^2 - \frac{\left( \delta \epsilon \right)^2}{4} \right) = \frac{\Rey}{2 \epsilon^2 \lambda} p(x_i,0) \left( (x - x_i)^2 - \frac{\left( \delta \epsilon \right)^2}{4} \right). \label{eqn:LubV}
\end{align}
Therefore, the dimensionless flux through each $i$-th branched channel is given by
\begin{align}
    Q_i^{\text{branch}} &= - \int_{x_i - \frac{\delta \epsilon}{2}}^{x_i + \frac{\delta \epsilon}{2}} v (x,0) \, \mathrm{d}x\\
    &= - \left[ \frac{\Rey}{2 \epsilon^2 \lambda}p(x_i,0) \int^{x_i + \frac{\delta \epsilon}{2}}_{x_i -\frac{\delta \epsilon}{2}}\left( (x - x_i)^2 - \frac{(\delta \epsilon)^2}{4} \right) \mathrm{d}x \right]\\
    &= \frac{\epsilon \delta^3 \Rey}{12 \lambda} p(x_i, 0). \label{eqn:dpdy}
\end{align}

We suppose that the total fluid flux through all the branched channels over $x \in [0,1]$ is $\mathcal{O}(1)$, so the fluid flux through each branched channel, $Q_i^{\text{branch}}$, is $\mathcal{O}(\epsilon)$. Therefore,
\begin{equation}
       Q_i^{\text{branch}} = \epsilon \kappa p(x_i,0),\label{eqn:DimlessFluxStrength}
\end{equation}
where we define
\begin{equation}
    \kappa = \frac{\delta^3 \Rey}{12 \lambda} = \mathcal{O}(1).
    \label{eqn:kappa}
\end{equation}

We now take the limit as $\delta \rightarrow 0$, with $\kappa = \mathcal{O}(1)$, so that the width of each branched channel vanishes. This allows us to consider the remaining problem on a simplified domain $(x,y) \in [0,1] \times [0, \gamma]$, where the branched channels are approximated by $2$--dimensional point-sinks in the bulk flow located along the bottom wall. Since each sink only draws in fluid from $y>0$, each has strength given by twice $Q_i^{\text{branch}}$ \citep{batchelor2000introduction}. 
Thus we replace \eqref{eqn:DimConsMass} with
\begin{equation}
\boldsymbol{\nabla} \cdot \boldsymbol{u} =- 2\epsilon \kappa \sum_{i=1}^N  p(x_i,0) \Delta \left(x-x_i, \,  y \right),
\label{eqn:ConsMassDeltaFunc}
\end{equation}
where $\Delta(\cdot,\cdot)$ is the 2-dimensional delta function, and we henceforth focus on the geometry shown in figure \ref{fig:PointSinkApproximation}.

\subsection{Inviscid approximation}

Since we assume that the Reynolds number is large, there \delete{will be}\add{exist} viscous boundary layers, of thickness $1/\sqrt{\Rey}$, on the top and bottom wall of the main channel. In addition, there \delete{will be}\add{is} a further layer, of thickness $\epsilon$, over which the effects due to fluid loss down the branched channels will be smoothed out.

Outside of the viscous boundary layers and away from the bottom wall, we simplify the governing equations~(\ref{eqn:DimConsMass}) and (\ref{eqn:DimNS}) by supposing that the flow is inviscid and irrotational. This leaves us with the system of equations,
\begin{align}
    \pd{{u}}{{x}} + \pd{{v}}{{y}} &= 0, \label{eqn:invConsMass}\\
    \pd{{v}}{{x}} - \pd{{u}}{{y}} &= 0, \label{eqn:invIrrot}\\
    \frac{1}{\epsilon^2} p + \frac{1}{2} |{\boldsymbol{u}}|^2 &= \text{constant}, \label{eqn:invBernoulli}
\end{align}
outside of the viscous boundary layers, where we have not included the right-hand side of \eqref{eqn:ConsMassDeltaFunc} since we are far away from the point-sinks. We note that, at leading order in $\epsilon$, the pressure, $p$, given by (\ref{eqn:invBernoulli}), is constant everywhere.

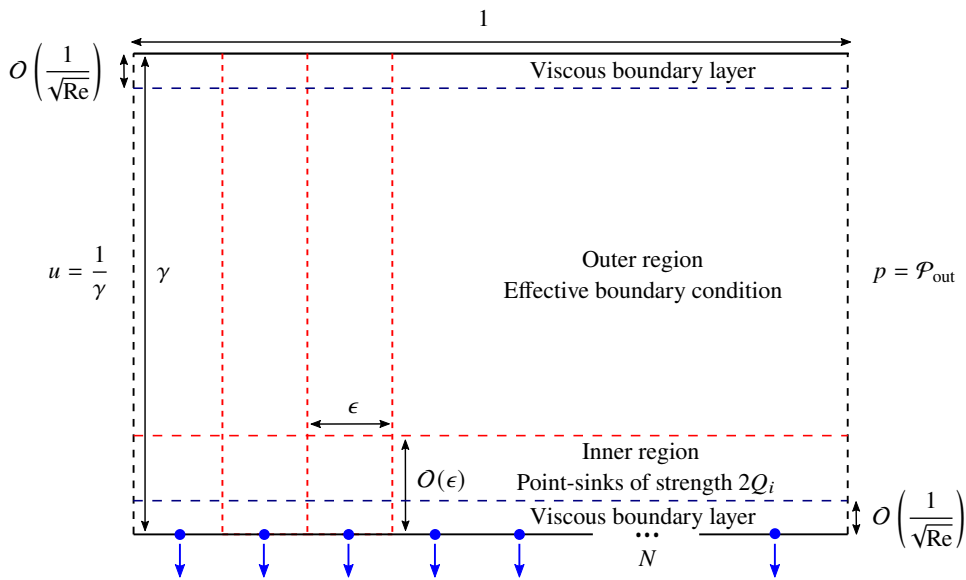
\begin{figure}
    \centering
    \vspace{5mm}
    \begin{overpic}[width=0.85\textwidth]{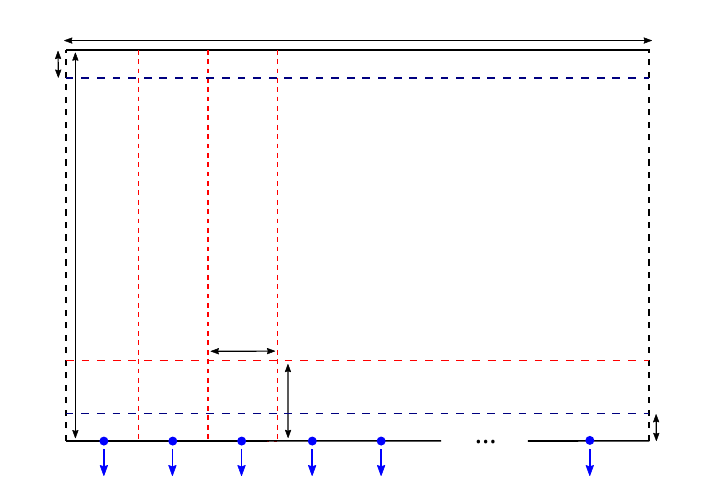}
    \put(34,22.5){$\epsilon$}
    \put(12,38){$\gamma$}
    \put(67.5,5){$N$}
    \put(49,67){$1$}
    \put(42,14){$\mathcal{O}(\epsilon)$}
    \put(94.5,38){$p=\mathcal{P}_\text{out}$}
    \put(-0.5,38){$u = \mathlarger{\f{1}{\gamma}}$}
    \put(61,39.25){Outer region}
    \put(52,35.75){Effective boundary condition}
    \put(61,17.25){Inner region}
    \put(53.5,13.75){Point-sinks of strength $2 Q_i$}
    \put(55,9.75){Viscous boundary layer}
    \put(55,61){Viscous boundary layer}
    \put(94.5,9.75){$\mathcal{O}\left(\mathlarger{\f{1}{\sqrt{\Rey}}}\right)$}
    \put(-5,61){$\mathcal{O}\left(\mathlarger{\f{1}{\sqrt{\Rey}}}\right)$}
    \end{overpic} 
    \vspace{3mm}
    \caption{Reduced dimensionless geometry, with point-sinks replacing each branched channel. Each point-sink has coordinates $(x_i, 0)$, where $x_i = (i-1/2) \epsilon$ for $i = 1, 2, \cdots, N$, and has strength $2 Q_i^{\text{branch}}$, where $Q_i^{\text{branch}}$ is the flux through a single channel, as in equation~(\ref{eqn:DimlessFluxStrength}), \citep{batchelor2000introduction}. The outer problem views the point-sinks as an effective boundary condition, capturing the overall average behaviour. Both boundary layers are indicated in the regime $\epsilon \gg 1/\sqrt{\Rey}$.}
    \label{fig:PointSinkApproximation}
\end{figure}

When considering the combination of the two boundary layers on the bottom wall, there are three possible regimes: (i) the viscous boundary layer is larger such that $\epsilon \ll 1/\sqrt{\Rey}$; (ii) the branched-channel-inducing boundary layer is larger, such that $\epsilon \gg 1/\sqrt{\Rey}$; (iii) the layers are of the same order, \add{i.e.,} $\epsilon = O(1/\sqrt{\Rey})$. We focus our attention on case (ii), as indicated in figure~\ref{fig:PointSinkApproximation}, so that the flow inside the $\epsilon$\add{-}boundary layer is also governed by (\ref{eqn:invConsMass})--(\ref{eqn:invBernoulli}), and we neglect the viscous boundary layer for simplicity, since the focus of our interest is in the effect of the branched channels.  The inviscid approximation of the flow everywhere reduces the complexity of the problem, and so we require fewer boundary conditions (see \S\ref{section:outer} and \S\ref{section:inner}). 

Across the $\epsilon$-boundary layer, the flow generated by the point-sinks will be smoothed out and our aim is to determine an effective boundary condition to be imposed on the outer flow.
We adopt the following solution procedure. We first partially solve for the outer flow. Scaling into the $\epsilon$-boundary layer, we solve for the flow close to the point-sinks. We then match between the two regions to find the effective boundary condition and a composite flow solution.

\section{Outer flow problem}

\label{section:outer}

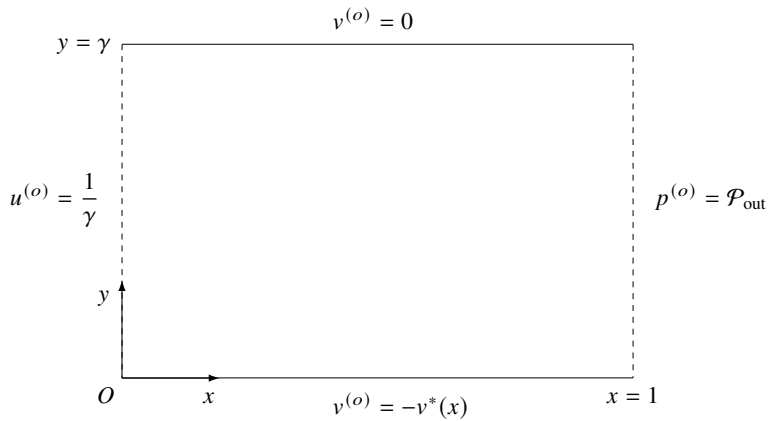
\begin{figure}
\centering
\vspace{4mm}
\begin{overpic}[width=0.8\textwidth]{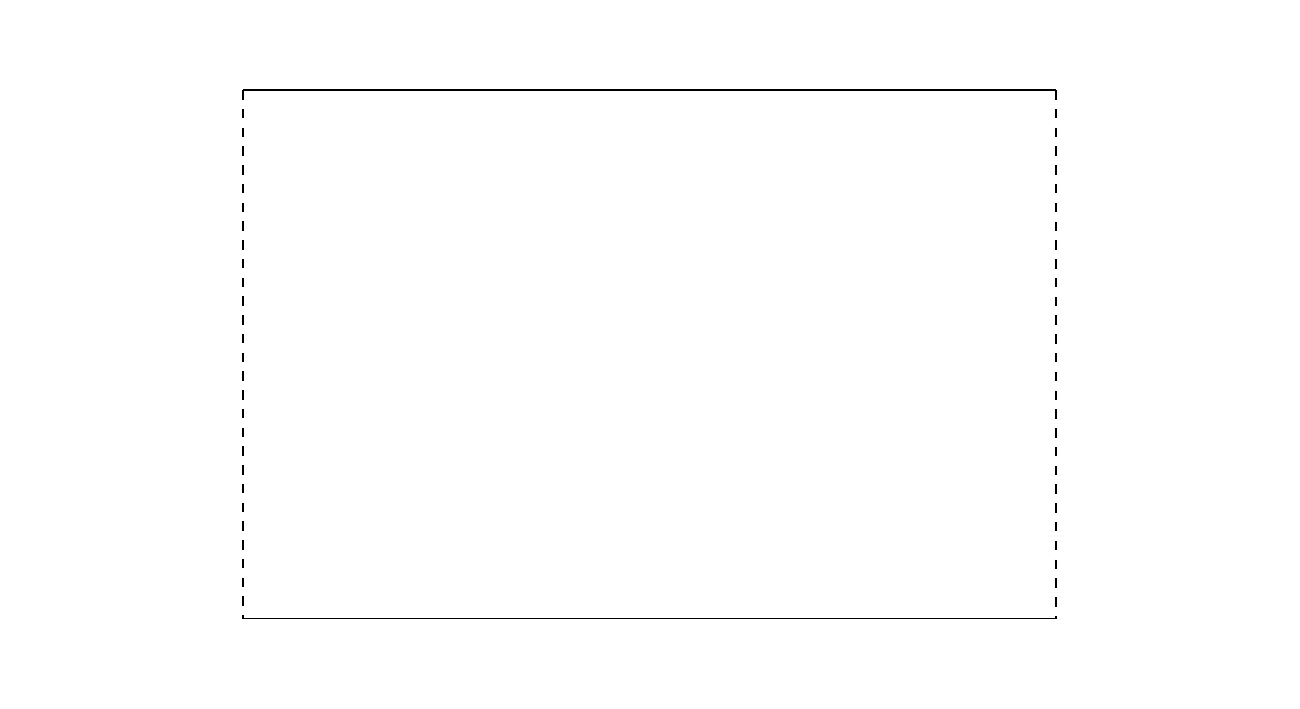}
    \put(10.7,47){$y=\gamma$}
    \put(78.1,3.9){$x=1$}
    \put(5,28){$u^{(o)} = \mathlarger{\f{1}{\gamma}}$}
    \put(84,28){$p^{(o)} = \mathcal{P}_\text{out}$}
    \put(44.5,49.5){$v^{(o)} = 0$}
    \put(44.5,2.5){$v^{(o)} = - v^*(x)$}
    %
    %
    \put(18.7,6.9){\vector(0,1){12}}
    \put(15.7,16.3){$y$}
    \put(18.7,6.9){\vector(1,0){12}}
    \put(28.7,3.9){$x$}
    \put(15.7,3.9){$O$}
    \end{overpic} 
    \vspace{2mm}
    \caption{Outer flow domain with boundary conditions, including the effective boundary condition, $v^{(o)} (x, 0) = - v^* (x)$.}
    \label{fig:OuterSinks}
\end{figure}

In the outer flow, ${\boldsymbol{u}}^{(o)} = \left( {u}^{(o)},{v}^{(o)} \right)$ and ${p}^{(o)}$ denote\delete{s} the outer velocity and pressure\add{,} respectively, which satisfy equations~(\ref{eqn:invConsMass})--(\ref{eqn:invBernoulli}). The boundary conditions for this outer problem (see figure~\ref{fig:OuterSinks}) become
\begin{align}
    u^{(o)} = \frac{1}{\gamma}, \quad &\text{on} \quad x = 0, \label{eqn:inletConst}\\
    v^{(o)} = 0, \quad &\text{on} \quad {y} = \gamma,\label{eqn:topwallnoflux}\\
    p^{(o)} = \mathcal{P}_\text{out}, \quad &\text{on} \quad {x} = 1, \label{eqn:outerPressureCond}\\
     v^{(o)}_0 = - v^*(x), \quad &\text{on} \quad y = 0,
    \label{eqn:OuterEffective}
\end{align}
where $v^*(x)$ is an unknown function to be determined via matching to the inner problem.

\subsection{Outer solution}

To solve for the leading-order outer solution, we consider an asymptotic expansion in $\epsilon$ of the form
\begin{equation}
    f^{(o)} (x,y) = f^{(o)}_0 (x,y) + \epsilon f^{(o)}_1 (x,y) + \cdots,
    \label{eqn:outerExp}
\end{equation}
for the dependent variables $u^{(o)}$, $v^{(o)}$\add{,} and $p^{(o)}$. Considering (\ref{eqn:invBernoulli}) with boundary condition~(\ref{eqn:outerPressureCond}), we find that the leading-order pressure is given by
\begin{equation}
    p^{(o)}_0 (x,y) = \mathcal{P}_\text{out},
    \label{eqn:LOOuterPressure}
\end{equation}
everywhere. 
Since the outer flow is irrotational, we introduce a velocity potential, $\phi^{(o)}$, such that
\abeqn{}{u^{(o)}_0 = \pd{\phi^{(o)}_0}{x}, \quad v^{(o)}_0 = \pd{\phi^{(o)}_0}{y}, \label{eqn:outerpotentialCR}}
which will be useful for matching with the inner problem. To solve for the leading-order outer velocities, $u^{(o)}_0$ and $v^{(o)}_0$, we first need to determine $v^*(x)$. This involves solving the inner problem and so we address this next, before returning to complete the solution in the outer flow in \S \ref{sec:outerpart2}.

\section{Inner flow problem}

\label{section:inner}

Having partially solved for the outer problem, we now scale into the inner region via the scaling
\begin{equation}
    y = \epsilon Y.
\end{equation}
In this region, the discrete nature of the point-sinks becomes apparent, illustrated in figure~\ref{fig:PointSinkApproximation}. We denote the inner velocity by $\boldsymbol{u}^{(i)} = \left( u^{(i)},v^{(i)} \right)$ and pressure $p^{(i)}$, and the scaled versions of (\ref{eqn:ConsMassDeltaFunc}), \eqref{eqn:invIrrot}, and (\ref{eqn:invBernoulli}) in the inner region are
\begin{align}
    \pd{{u}^{(i)}}{x} + \frac{1}{\epsilon} \pd{v^{(i)}}{Y} &= - 2\epsilon \kappa \sum_{i=1}^N  p(x_i,0) \Delta \left(x-x_i, \,  \epsilon Y \right) \\
&= - 2 \kappa \sum_{i=1}^N  p(x_i,0) \Delta \left(x-x_i, \,  Y \right)\add{,}
\label{eqn:IntConsMassQ}\\
    \pd{v^{(i)}}{{x}} - \f{1}{\epsilon} \pd{u^{(i)}}{Y} &= 0, \label{eqn:CurlVelocity}\\
    \frac{1}{\epsilon^2} p^{(i)} + \frac{1}{2} \left( \left(u^{(i)} \right)^2 + \left( v^{(i)} \right)^2 \right) &= \text{constant} \label{eqn:BLayerBernoulliQ},
\end{align}
where we have used the property that $\Delta (\alpha x, \beta y) = \Delta (x, y) / \alpha \beta$.
The inner region boundary conditions are
\begin{align}
    u^{(i)} = \f{1}{\gamma}, \quad &\text{on} \quad {x} = 0, \label{eqn:InletHoriz}\\
    v^{(i)} = 0, \quad &\text{on} \quad Y = 0,\\
    p^{(i)} = {\mathcal{P}}_\text{out}, \quad &\text{on} \quad x = 1, \label{eqn:OutPressure}
\end{align}
with matching conditions to the outer region,
\begin{align}
    u^{(i)} (x, Y) \rightarrow u^{(o)}(x, 0), \quad &\text{as} \quad Y \rightarrow + \infty, \label{eqn:MatchingUInner}\\
    v^{(i)} (x, Y) \rightarrow v^{(o)} (x, 0), \quad &\text{as} \quad Y \rightarrow + \infty. \label{eqn:MatchingVInner}
\end{align}

As in the outer, since the problem is irrotational, we choose to express the inner problem in terms of a velocity potential, $\phi^{(i)}$, defined by the ansatz,
\abeqn{}{u^{(i)} = u^{(o)}_0 (x, \epsilon Y) + \epsilon \pd{\phi^{(i)}}{x}, \quad v^{(i)} = v^{(o)}_0 (x, \epsilon Y) + \frac{\partial \phi^{(i)}}{\partial Y}. \label{eqn:CauchyRiemann}} 
We take this form so that we automatically match to the leading-order outer velocities, $u^{(o)}_0$ and $v^{(o)}_0$, where any variations are introduced via $\phi^{(i)}$. Substituting \eqref{eqn:CauchyRiemann} into equation (\ref{eqn:IntConsMassQ}) gives
\begin{equation}
    \epsilon \frac{\partial^2 \phi^{(i)}}{\partial x^2} + \frac{1}{\epsilon} \frac{\partial^2 \phi^{(i)}}{\partial Y^2} =  - 2 \kappa \sum_{i=1}^N p^{(i)}(x_i, 0) \Delta \left( x-x_i, \, Y \right),
    \label{eqn:BulkConsMassSummation}
\end{equation}
since
\begin{equation}
    \pd{{u}^{(o)}_0}{x} (x, \epsilon Y) + \frac{1}{\epsilon} \pd{{v}^{(o)}_0}{Y} (x, \epsilon Y) = 0,
\end{equation}
from \eqref{eqn:invConsMass} in the outer problem, and equation~(\ref{eqn:BLayerBernoulliQ}) becomes
\begin{equation}
   \frac{1}{\epsilon^2} p^{(i)} + \frac{1}{2} \left[ \left( u^{(o)}_0 (x, \epsilon Y) + \epsilon \pd{\phi^{(i)}}{x} \right)^2 + \left( v^{(o)}_0 (x, \epsilon Y) + \frac{\partial \phi^{(i)}}{\partial Y} \right)^2 \right] = \text{constant}.
    \label{eqn:BernouliInnerPhi}
\end{equation}
The boundary conditions~(\ref{eqn:InletHoriz})--(\ref{eqn:OutPressure}) become\delete{,}
\begin{align}
    \pd{\phi^{(i)}}{x} = 0, \quad &\text{on} \quad {x} = 0, \label{eqn:InletHorizPhi}\\
    \pd{\phi^{(i)}}{Y} = 0, \quad &\text{on} \quad Y = 0, \label{eqn:InnerWallCond}\\
    p^{(i)} = {\mathcal{P}}_\text{out}, \quad &\text{on} \quad x = 1.\label{eqn:OutPressurePhi}
\end{align}

Before proceeding, we comment on the matching conditions~(\ref{eqn:MatchingUInner}) and (\ref{eqn:MatchingVInner}) between the inner and outer solutions. Given our \change{anzatz}{ansatz} for $u^{(i)}$ and $v^{(i)}$ \delete{given} in \eqref{eqn:CauchyRiemann}, the leading-order matching is automatically satisfied. We \change{will also need to use}{also follow \cite{van1975perturbation} to establish an additional matching condition between the inner and outer solutions, using}
\begin{equation}
   \delete{1 \mathrm{to} ( 2 \mathrm{ti} ) = 2 \mathrm{ti} (1 \mathrm{to}).} \add{ 1 \, \text{term outer} \, ( 2 \, \text{term inner} ) = 2 \, \text{term inner} \, (1 \, \text{term outer} ).}
\end{equation}
\delete{We follow Van Dyke (1975) to establish this condition and, w}Writing $\phi^{(i)} \sim \phi^{(i)}_1 + \epsilon\phi^{(i)}_2 + \cdots$, \change{we find that}{this matching condition becomes} 
\begin{equation}
    \lim_{Y \rightarrow + \infty} \pd{\phi^{(i)}_1}{Y} (x,Y) = \pd{\phi^{(o)}_0}{y} (x, 0) = -v^*(x).
    \label{eqn:matchingphi1Y}
\end{equation}
    
\subsection{Method of multiple scales}

The domain within the inner region is made up of a repeating structure of length $\epsilon$, surrounding each point-sink. We describe each part of the repeating structure as a cell. We assume that the flow is slowly varying over each cell. We introduce a microscale variable, $X$, by letting $x = \epsilon X$, to capture both the behaviour over the \change{device}{domain} length and \add{over the length of} each cell.
We treat both the long scale, $x$, and short scale, $X$, as independent variables and assume that all \textit{dependent} variables depend on $x, X$, and $Y$ in the inner region. Since we treat each of these variables as independent, spatial derivatives in $x$ transform as
\begin{equation}
    \pd{}{x} \mapsto \pd{}{x} + \f{1}{\epsilon} \pd{}{X}.
\end{equation}
The sink in a particular cell is located at $X=0$. The addition of the \delete{multiscale assumption}\ownadd{microscale variable} also gives us an additional degree of freedom. We remove this by imposing periodicity on the boundary walls such that
\begin{equation}
    \pd{\phi^{(i)}}{X} \bigg \rvert_{X=-\frac{1}{2}} = \pd{\phi^{(i)}}{X} \bigg \rvert_{X = \frac{1}{2}} = 0.
    \label{eqn:InnerPeriodicCondition}
\end{equation}
This retains the slow variation of $u^{(i)}$ over each cell, where (\ref{eqn:CauchyRiemann}a) becomes
\begin{align}
    u^{(i)} = u^{(o)}_0 (x, \epsilon Y) + \pd{\phi^{(i)}}{X}+ \epsilon \pd{\phi^{(i)}}{x}.
    \label{eqn:uPhiEqn}
\end{align}

We solve for $\phi^{(i)}$ in each periodic cell as a point-sink cell problem in a periodic semi-infinite half strip, and match to the outer solution. The inner bulk equation~(\ref{eqn:BulkConsMassSummation}) in the cell problem becomes
\begin{equation}
    \frac{\partial^2 \phi^{(i)}}{\partial X^2} + \frac{\partial^2 \phi^{(i)}}{\partial Y^2} + 2 \epsilon \frac{\partial^2 \phi^{(i)}}{\partial x \partial X} + \epsilon^2 \frac{\partial^2 \phi^{(i)}}{\partial x^2} =  - 2 \kappa p^{(i)}(x_i, 0, 0) \Delta(X , Y), \label{eqn:Potentialhomogenised}
\end{equation}
where we have used the fact that $\epsilon \Delta(x-x_i,Y)=\Delta(X,Y)$.

\subsection{Inner solution}

We consider an asymptotic expansion of the inner pressure to be of the form
\begin{equation}   p^{(i)} (x, X, Y) = p^{(i)}_0 (x, X, Y) + \epsilon p^{(i)}_1 (x, X, Y) + \cdots. \label{eqn:innerExp}
\end{equation}

\begin{figure}
\centering
\begin{overpic}[width=0.9\textwidth]{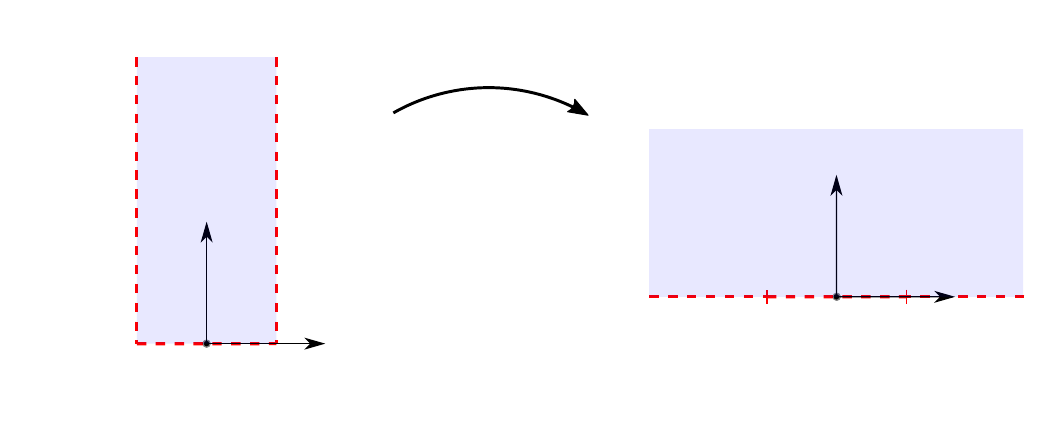}
    \put(20.5,10.5){$O$}
    \put(21,19){$Y$}
    \put(29.25,10.5){$X$}
    \put(26,4.5){$\mathlarger{\frac{1}{2}}$}
    \put(9.25,4.5){$\mathlarger{-\frac{1}{2}}$}
    \put(15.25,5){$\phi^{(i)}_{1Y} = 0$}
    \put(28,23){$\phi^{(i)}_{1X} = 0$}
    \put(2,23){$\phi^{(i)}_{1X} = 0$}
    \put(39.75,35.5){$\zeta = \sin{(\pi Z)}$}
    \put(79.75,14.75){$O$}
    \put(88,14.75){$\Re(\zeta)$}
    \put(80,23.5){$\Im(\zeta)$}
    \put(84.6,14.75){$1$}
    \put(71,14.75){$-1$}
    \put(62,9){$\phi^{(i)}_{1 \eta} = 0$}
    \put(74.25,9){$\phi^{(i)}_{1 \eta} = 0$}
    \put(86.5,9){$\phi^{(i)}_{1 \eta} = 0$}
    \end{overpic} 
    \caption{Conformal map of the semi-infinite half strip inner region to the positive imaginary half plane via the conformal map $\zeta = \sin{(\pi Z)}$.}
    \label{fig:conformalmap}
\end{figure}
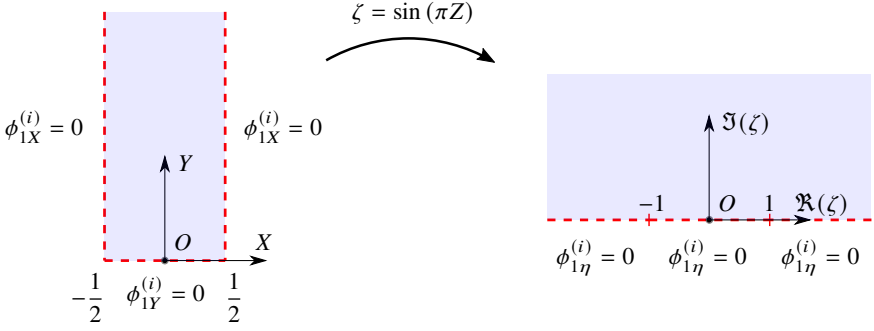

We first solve for the leading-order pressure in the inner region. The leading-order version of (\ref{eqn:BernouliInnerPhi}) implies that $p^{(i)}_0$ is a constant and so, matching with the solution in the outer flow given by (\ref{eqn:LOOuterPressure}), we find that
\begin{equation}
    p^{(i)}_0 (x, X, Y) = \mathcal{P}_\text{out}.
    \label{eqn:LOInnerPressure}
\end{equation}
Thus, the leading-order pressure is constant everywhere in the flow. The problem for $\phi^{(i)}_1$ is given by\delete{,}
\begin{equation}
    \frac{\partial^2 \phi^{(i)}_1}{\partial X^2} + \frac{\partial^2 \phi^{(i)}_1}{\partial Y^2} = - 2 \kappa \mathcal{P}_\text{out} \Delta(X , Y), \label{eqn:FirstOrderPhiConsMass}
\end{equation}
with boundary and matching conditions\delete{,}
\begin{align}
    \pd{\phi^{(i)}_1}{X} = 0, \quad &\text{on} \quad X = \pm \frac{1}{2}, \; Y \in (0,\infty), \label{eqn:FirstOrderPhiPeriodic}\\
    \pd{\phi^{(i)}_1}{Y} = 0, \quad &\text{on} \quad Y=0, \; X \in \left[-\f{1}{2}, \f{1}{2} \right],\\
    \pd{\phi^{(i)}_1}{Y} \rightarrow -v^*(x), \quad &\text{as} \quad Y \rightarrow +\infty, \; X \in \left[-\f{1}{2}, \f{1}{2} \right]. \label{eqn:FirstOrderPhiInfCond}
\end{align}
We use complex variable theory to find the solution for $\phi^{(i)}_1$ in the semi-infinite periodic strip with a point-sink at the origin of strength $2 \kappa \mathcal{P}_\text{out}$. Using the holomorphic conformal mapping,
\begin{equation}
    \zeta = \xi + \mathrm{i} \eta = \sin{(\pi Z)},
\end{equation}
where $Z = X + \mathrm{i} Y$, the semi-infinite periodic strip domain is mapped to a semi-infinite half plane, $\eta = \Im(\zeta) \geq 0$, as indicated in figure~\ref{fig:conformalmap}, where we denote $\Re$ and $\Im$ as the real and imaginary parts, respectively. Laplace's equation still holds in the transformed domain, however the periodic conditions become no-flux conditions on the half-plane boundary.

The problem in the conformally mapped space is thus
\begin{equation}
    \nabla^2_\zeta \phi^{(i)}_1 = - 2 \kappa\mathcal{P}_\text{out} \Delta{(\zeta)}, \label{eqn:ConformalProblem1}
\end{equation}
with boundary and matching conditions
\begin{align}
    \pd{\phi^{(i)}_1}{\eta} = 0, \quad &\text{on} \quad \Im (\zeta) = 0,\\
    \pd{\phi^{(i)}_{1}}{\eta} \rightarrow - v^*(x), \quad &\text{as} \quad \Im (\zeta) \rightarrow \infty. \label{eqn:ConformalProblem3}
\end{align}

The solution to (\ref{eqn:ConformalProblem1})--(\ref{eqn:ConformalProblem3}) is the well-known solution to Laplace's equation with a point-sink/source,
\begin{equation}
    \phi_1^{(i)}(x,\zeta) = C(x)- \f{\kappa \mathcal{P}_\text{out}}{\pi} \log(\zeta),
\end{equation}
 where $C(x)$ is a function of integration. Transforming back into $(x,X,Y)$ coordinates, we have
\begin{equation}
    \phi_1^{(i)}(x,X,Y) = C(x)- \f{\kappa \mathcal{P}_\text{out}}{\pi} \Re[ \log ( \sin{ (\pi (X + \mathrm{i} Y) } ) ].
    \label{eqn:InnerPhi1}
\end{equation}
The function of integration, $C(x)$, may be taken to be zero without loss of generality. Therefore, since
\begin{align}
    \pd{\phi^{(i)}_1}{X} &= - \kappa \mathcal{P}_\text{out} \Re \left[ \cot{ \left( \pi \left( X + \mathrm{i} Y \right) \right)} \right]\\
    &= -  
    \frac{ \kappa \mathcal{P}_\text{out}
    \sin{\left(\pi X \right)} \cos{\left( \pi X  \right)}
    }{
    \cos^2{\left(\pi X  \right)} \sinh^2{\left(\pi Y \right)} + \sin^2{\left(\pi X \right)} \cosh^2{\left(\pi Y \right)}
    },
    \label{eqn:InnerDiffX}
\end{align}
and
\begin{align}
    \pd{\phi^{(i)}_1}{Y} &= \kappa \mathcal{P}_\text{out} \Im \left[ \cot{ \left( \pi \left( X + \mathrm{i} Y \right) \right)} \right]\\
    &= - \frac{ \kappa \mathcal{P}_\text{out}
    \cosh{\left(\pi Y \right)} \sinh{\left( \pi Y \right)}
    }{
    \cos^2{\left(\pi X \right)} \sinh^2{\left(\pi Y \right)} + \sin^2{\left(\pi X \right)} \cosh^2{\left(\pi Y \right)}
    },
    \label{eqn:v1Inner}
\end{align}
the leading-order $x$-component of the inner velocity is given by\delete{,}
\begin{equation}
    u^{(i)}_0 (x, X, Y) = u^{(o)}_0 (x, 0)  -  
    \frac{ \kappa \mathcal{P}_\text{out}
    \sin{\left(\pi X \right)} \cos{\left( \pi X  \right)}
    }{
    \cos^2{\left(\pi X  \right)} \sinh^2{\left(\pi Y \right)} + \sin^2{\left(\pi X \right)} \cosh^2{\left(\pi Y \right)}
    },
    \label{eqn:uLOInner}
\end{equation}
and\delete{,} the leading order $y$-component of the inner velocity is given by
\begin{equation}
    v^{(i)}_0 (x, X, Y) = v^{(o)}_0 (x, 0) - \frac{ \kappa \mathcal{P}_\text{out}
    \cosh{\left(\pi Y \right)} \sinh{\left( \pi Y \right)}
    }{
    \cos^2{\left(\pi X \right)} \sinh^2{\left(\pi Y \right)} + \sin^2{\left(\pi X \right)} \cosh^2{\left(\pi Y \right)}
    }.
    \label{eqn:vLOInner}
\end{equation}
Having found the solution in the inner region, we calculate that, as $Y \rightarrow \infty$, 
\begin{align}
    \pd{\phi^{(i)}_1}{Y} &\rightarrow - \kappa \mathcal{P}_\text{out} \frac{ 
    e^{\pi Y} e^{\pi Y} 
    }{
    \cos^2{\left(\pi X \right)} e^{2 \pi Y} + \sin^2{\left(\pi X \right)} e^{2 \pi Y}
    }
    = - \kappa \mathcal{P}_\text{out}.
\end{align}
Therefore, via the matching condition~\eqref{eqn:FirstOrderPhiInfCond}, we find that
\begin{equation}
v^* (x) = \kappa \mathcal{P}_\text{out},
\label{eqn:effectiveBC}
\end{equation}
and thus 
\begin{equation}
    v^{(o)}_0 (x, 0) =-\kappa \mathcal{P}_\text{out}.\label{eqn:newv}
\end{equation}
Thus, the boundary layer smooths out the variation caused by the discrete sinks\add{,} and the outer flow simply experiences a spatially uniform flow of liquid out through  the ``effective" bottom of the device.

\section{Outer solution}

\label{sec:outerpart2}

Now that we have found $v^*(x) = \kappa \mathcal{P}_\text{out}$, we return to the outer problem to find the leading-order outer velocities $u^{(o)}_0$ and $v^{(o)}_0$. Solving equations~(\ref{eqn:invConsMass}) and (\ref{eqn:invIrrot}), with boundary conditions \eqref{eqn:inletConst}, \eqref{eqn:topwallnoflux}, \eqref{eqn:OuterEffective}\add{,} and \eqref{eqn:newv}, we find the simple solution
\abeqn{}{
    u^{(o)}_0 (x,y) = \frac{1 - \kappa \mathcal{P}_\text{out} x}{\gamma}, 
    \qquad v^{(o)}_0 (x,y) = -\kappa \mathcal{P}_\text{out}\left(1 - \frac{y}{\gamma}\right).
    \label{eqn:LOvOuter}
}
We see that $u_0^{(o)}$is independent of $y$ and so $u^{(o)}_0 (x, \epsilon Y)=u^{(o)}_0 (x, 0)=u^{(o)}_0 (x, y)$.

\section{Composite solution and outflow flux}

\label{section:CompSolution}

Using \eqref{eqn:uLOInner} and (\ref{eqn:LOvOuter}a), we write the leading-order composite solution for $u$ as
\begin{equation}
    \hspace{-1mm} u_c (x, y) = \frac{1 - \kappa \mathcal{P}_\text{out} x}{\gamma} -  
    \frac{ \kappa \mathcal{P}_\text{out}
    \sin{\left[\frac{\pi}{\epsilon} \left( x - \frac{\epsilon}{2} \right) \right]} \cos{\left[ \frac{\pi}{\epsilon} \left( x - \frac{\epsilon}{2} \right) \right]}
    }{
    \cos^2{\left[\frac{\pi}{\epsilon} \left( x - \frac{\epsilon}{2} \right) \right]} \sinh^2{\left[\frac{\pi y}{\epsilon} \right]} + \sin^2{\left[\frac{\pi}{\epsilon} \left( x - \frac{\epsilon}{2} \right) \right]} \cosh^2{\left[\frac{\pi y}{\epsilon} \right]}
    },
    \label{eqn:compU}
\end{equation}
where we have taken care to correctly locate the sinks by setting $X=(x-\epsilon/2)/\epsilon$. Similarly, using \eqref{eqn:vLOInner} and (\ref{eqn:LOvOuter}b), we find the leading-order composite solution for $v$ is
\begin{equation}
    \hspace{-1.5mm} v_c (x,y) = \kappa \mathcal{P}_\text{out} \left( \frac{y}{\gamma} -\frac{
    \cosh{\left[\frac{\pi y}{\epsilon} \right]} \sinh{\left[ \frac{\pi y}{\epsilon} \right]}
    }{
    \cos^2{\left[\frac{\pi}{\epsilon} \left( x - \frac{\epsilon}{2} \right) \right]} \sinh^2{\left[\frac{\pi y}{\epsilon} \right]} + \sin^2{\left[\frac{\pi}{\epsilon} \left( x - \frac{\epsilon}{2} \right) \right]} \cosh^2{\left[\frac{\pi y}{\epsilon} \right]}
    } \right).
    \label{eqn:vComposite}
\end{equation} 
These explicit formulae for the velocities bypass the need to run computationally challenging numerical simulations. 

We calculate the total flux, $Q_T$, through the lower boundary, using
\begin{equation}
    Q_T = - \int^1_0 v^*(x) \, \mathrm{d}x = \frac{\delta^3 \Rey}{12 \lambda} \mathcal{P}_\text{out}, \label{eqn:WallLOFlux}
\end{equation}
while the flux through the main outlet is $1 - Q_T$. Hence, the dimensional flux, $\hat{Q}_T$, is
\begin{equation}
    \hat{Q}_T= \frac{ h_2^3 N}{12 L_1 \mu } \hat{\mathcal{P}}_\text{out}.
\end{equation}
Thus we see that the flux out through the bottom boundary scales in the obvious way --- $N$ times the flux out of a single channel in this limit in which the pressure is constant.

\section{Flow results}
\label{section:FlowResult}

To validate our asymptotic prediction of the fluid flow and effective boundary condition, we compare with numerical solutions of the flow in the full branched domain.

\subsection{Numerical results}

\label{section:FullNumerical}

\begin{figure}
    \centering
    \begin{overpic}[width=0.8\textwidth,tics=10]
    {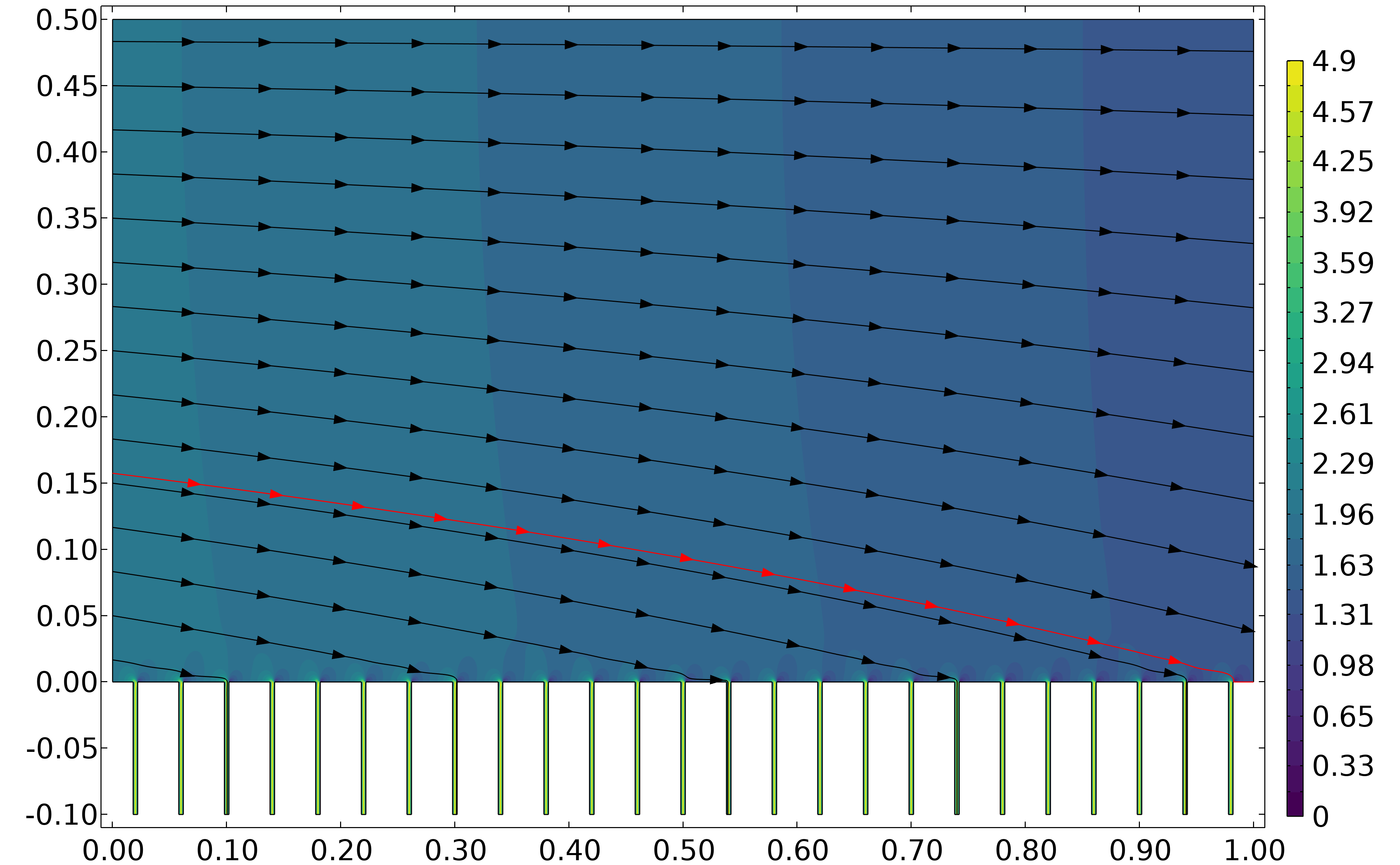}
    \put(48,-4){$x$}
    \put(-3,35){$y$}
    \put(101,35){$|\boldsymbol{u}|$}
    \end{overpic}
    \vspace{3mm}
\caption{Numerical solution for the magnitude of the flow velocity, $|\boldsymbol{u}|$, solved via the Navier--Stokes equations \eqref{eqn:DimConsMass}--\eqref{eqn:DimNS}. We apply a slip condition on the main channel walls and no-slip on the branched channel walls. Here, $\mathcal{P}_\text{out} = 0.4$, $\Rey = 1000$, $\epsilon = 0.04$, $\delta = 0.1$, $\lambda = 0.1$ and $\gamma = 0.5$. The black lines indicate streamlines and the red line indicates the dividing streamline.}
\label{fig:FullModel_Flow}
\end{figure}

We carry out numerical solutions in \texttt{COMSOL} in which we capture the nature of the high-Reynolds\add{-}number flow in the main channel and the \change{Poiseulle}{Poiseuille} flow in the branched channels. Since, in the regime of interest, the viscous boundary layers are much thinner than the $\epsilon$-layers and not a focus of this study, we impose free slip on the horizontal surfaces located at $y=0$ and $y=\gamma$ to ensure that the flow in the main channel emulates inviscid flow.
We illustrate the solution structure and the relevant result comparisons in Appendix \ref{appB}.  Thus, we solve  (\ref{eqn:DimConsMass})--(\ref{eqn:DimlessCentreBranch}), with \eqref{eqn:dimlessInletv0} removed and \eqref{eqn:DimlessSlip} partially replaced by the free-slip condition on the main channel walls. 
We refine the mesh such that the total flux through the thin branched channels converges to a constant value for any smaller mesh size. We use a once refined free triangular, extremely fine mesh in the outer domain and two mapped distribution meshes along $y=0$ and the walls of the branched channels. This achieves a convergent solution for the flow through the small branched channels, stable for any smaller mesh size.
\begin{figure}
    \centering
    \vspace{5mm}
    \begin{overpic}[width=0.43\textwidth,tics=10]
    {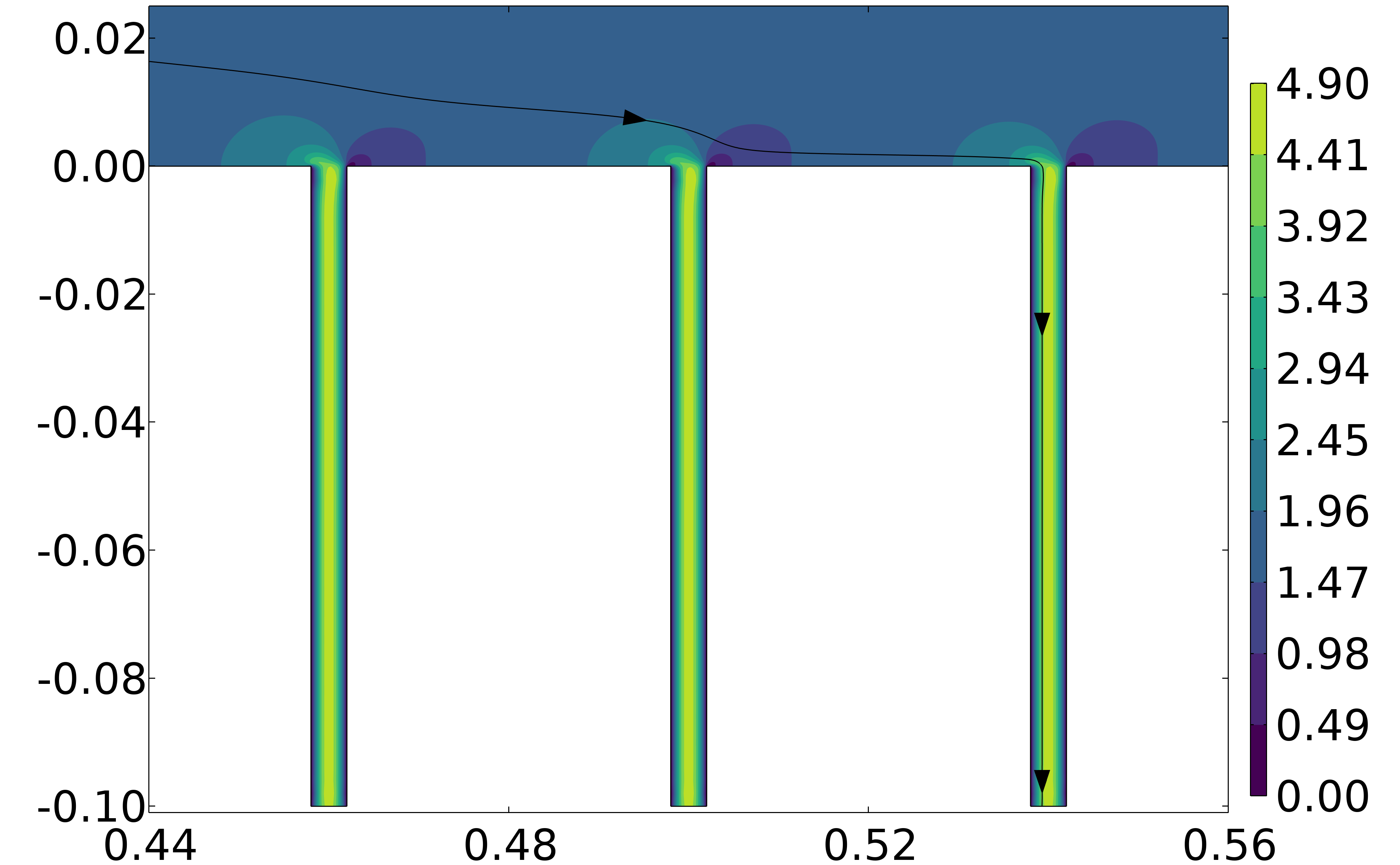}
    \put(47.75,-5.5){$x$}
    \put(-2.5,30){$y$}
    \put(100,30){$|\boldsymbol{u}|$}
    \put(46.5,67){(a)}
    \end{overpic}
    \hspace{10mm}
    \begin{overpic}[width=0.43\textwidth,tics=10]
    {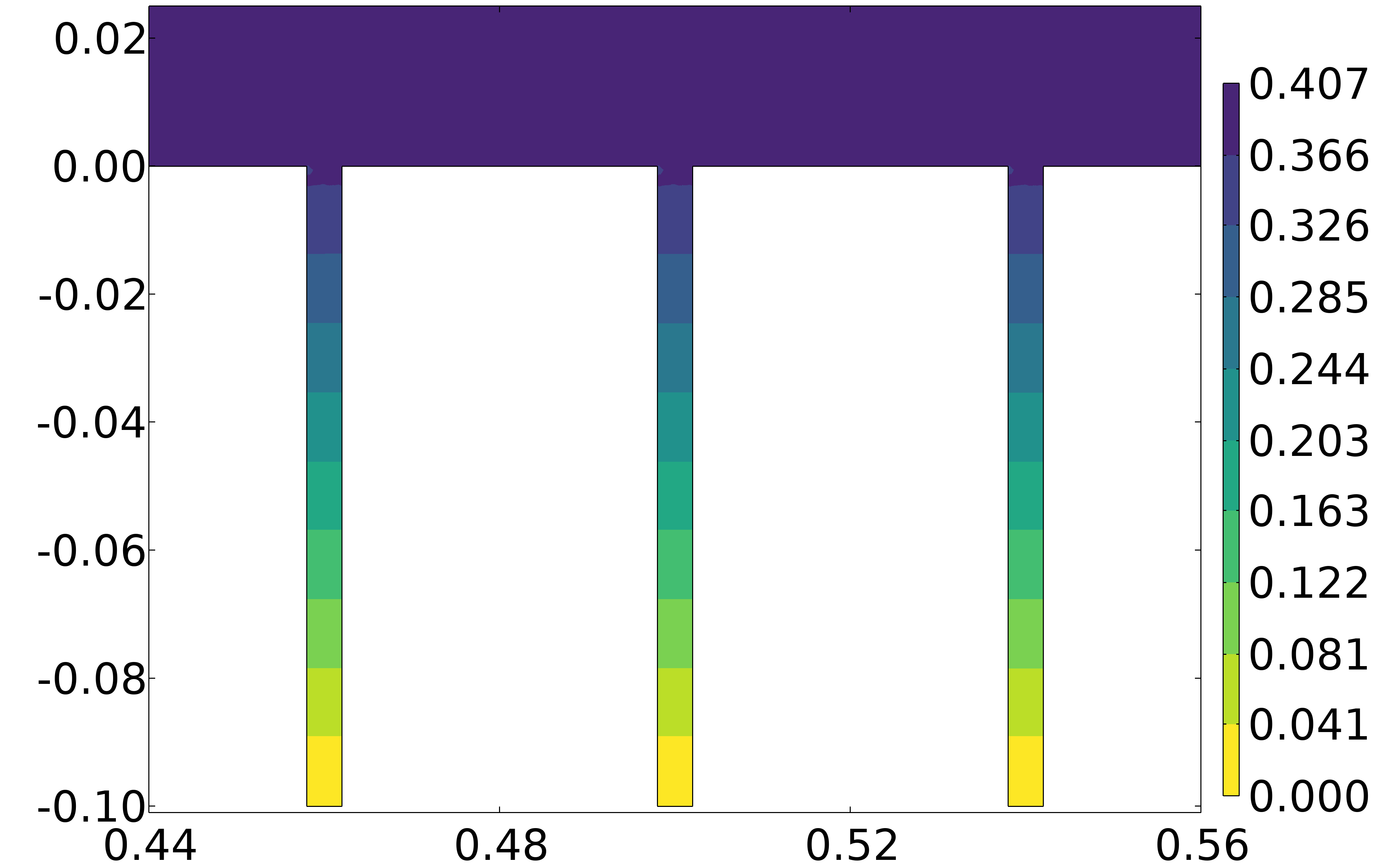}
    \put(47.75,-5.5){$x$}
    \put(-2.5,30){$y$}
    \put(100,30){$p$}
    \put(46.5,67){(b)}
    \end{overpic}
    \vspace{4mm}
\caption{Numerical solutions, zoomed \change{into}{in to} individual branched channels, for (a) the magnitude of the flow velocity, $|\boldsymbol{u}|$ and (b) pressure, $p$, from figures~\ref{fig:FullModel_Flow} and \ref{fig:FullModel_Pressure}\add{,} respectively, with the full colour range for $p$. Here, $\mathcal{P}_\text{out} = 0.4$, $\Rey = 1000$, $\epsilon = 0.04$, $\delta = 0.1$, $\lambda = 0.1$ and $\gamma = 0.5$. The black arrowed line indicates a particular streamline.}
\label{fig:Zoomed}
\end{figure}

\begin{figure}
    \centering
    \begin{overpic}[width=0.8\textwidth,tics=10]
    {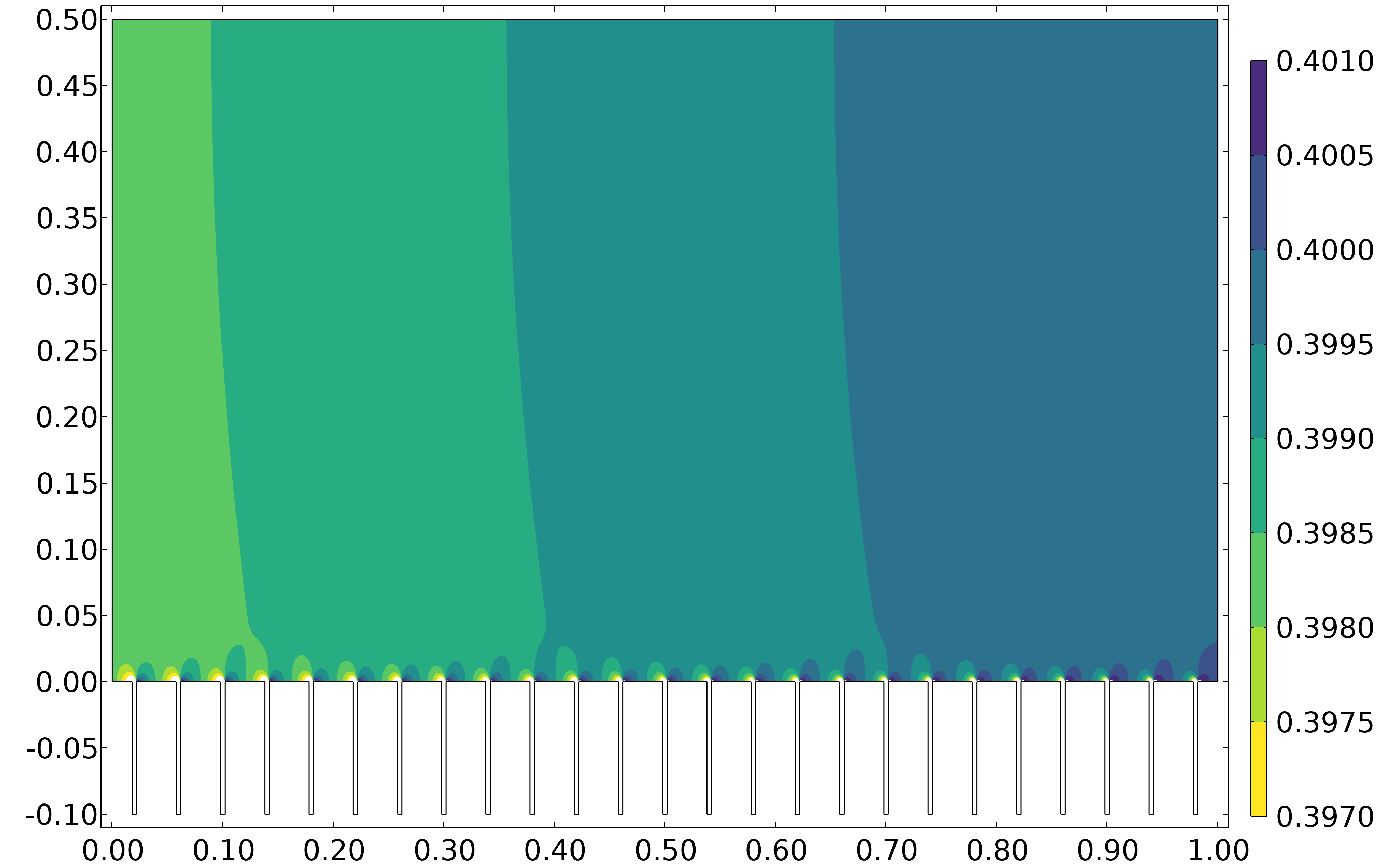}
    \put(48,-4){$x$}
    \put(-3,35){$y$}
    \put(101,35){$p$}
    \end{overpic}
    \vspace{3mm}
\caption{Numerical solution for the pressure, $p$, corresponding to figure~\ref{fig:FullModel_Flow}, solved via the Navier--Stokes equations \eqref{eqn:DimConsMass}--\eqref{eqn:DimNS}. Here, $\mathcal{P}_\text{out} = 0.4$, $\Rey = 1000$, $\epsilon = 0.04$, $\delta = 0.1$, $\lambda = 0.1$ and $\gamma = 0.5$. The colour range for the pressure is restricted to $[0.397,0.401]$, and white otherwise, to highlight the asymptotic observation of constant pressure to leading-order in $\epsilon$ over the main channel. The branched channels are shown to illustrate their location.}
\label{fig:FullModel_Pressure}
\end{figure}

\subsection{Asymptotic and numerical comparison}

We show the magnitude of the flow velocity resulting from our simulations and the streamlines of the flow, for an example set of parameters in which $\mathcal{P}_{\textrm{out}}=0.4$, $\textrm{Re}=1000$, $\epsilon=0.04$, $\delta=0.1$, $\lambda=0.1$, and $\gamma=0.5$, in figure~\ref{fig:FullModel_Flow}. We note that these values are on the edge of the asymptotic regime. However, this is the largest Reynolds number that we can achieve in COMSOL given the geometric complexities we are considering, but we \delete{will} also explore $\epsilon = 0.1$ (which has fewer channels than in the Beko PLC device) which sits more squarely in this asymptotic regime. We see that the liquid flows across from left to right and down, as expected, with a dividing streamline separating the flow that exits through the branched channels from the flow that leaves through the main channel outlet. We see in figure~\ref{fig:Zoomed}(a) that the flow velocity is largest in the branched channels. We calculate the total flux through each of the branches, $\sum_i Q_i^\text{branch}$ to be $0.315$ which corresponds to two times the height of the dividing streamline at $x=0$ --- this is because the inlet flux is $q=1$ and the height of the main channel is $\gamma = 0.5$. For the same parameter values, the asymptotic prediction~(\ref{eqn:WallLOFlux}) gives $Q =  1/3$, which is an $\mathcal{O}(\epsilon)$ difference.

We show the corresponding pressure in figure~\ref{fig:FullModel_Pressure}. We see that the pressure is almost constant in the main channel, and then decreases to zero along the branched channels (see figure \ref{fig:Zoomed}(b)). Maybe surprisingly at a first glance, we see that the pressure increases from left to right. The simulations  compare well with the leading-order asymptotic prediction, $p=\mathcal{P}_\text{out}$, with deviations of $\mathcal{O}(\epsilon)$ over the domain.

\begin{figure}
\centering
\vspace{0mm}
\begin{tabular}{ccc}
 & \quad \; (a) & \quad (b) \\[5mm]
    &  
    \begin{overpic}[width=0.4\textwidth,tics=10]
    {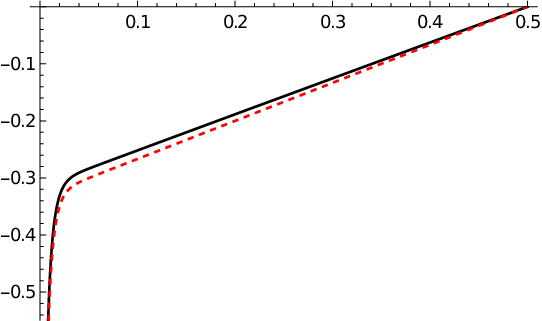}
    \put(51,62){$y$}
    \put(6,62){$v$}
    \put(3,22.8){\line(1,0){6}}
    \put(-17,21.4){$v^* = \mathlarger{\frac{1}{3}}$}
    \end{overpic} 
    &
    \begin{overpic}[width=0.42\textwidth,tics=10]
    {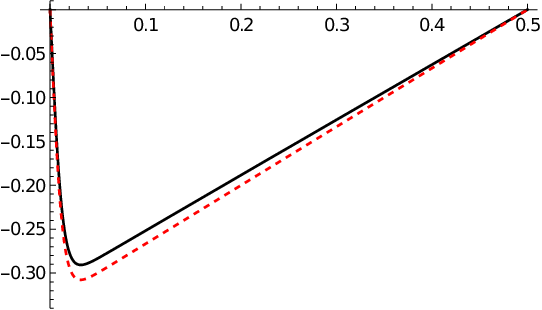}
    \put(52,59){$y$}
    \put(8,60){$v$}
    \put(5,1.5){\line(1,0){6}}
    \put(-17,0){$v^* = \mathlarger{\frac{1}{3}}$}
    \put(75,5){$\boxed{\epsilon = 0.04}$}
    \end{overpic}
\end{tabular}
\vspace{3mm}
\caption{Comparison of the leading-order asymptotic composite solution~\eqref{eqn:vComposite} in red dashed lines, with the numerical solution in solid black lines, for the $y$-component of velocity, $v$. The solutions are taken at (a) $x = x_i$, the centre of the branched channel/point-sink and (b) $x = x_i + \epsilon/2$, the right-hand edge of the periodic unit cell. In this example, we consider the centremost cell, taking $x_i = 0.5$, and vary the cell width. Here, $\mathcal{P}_\text{out} = 0.4$, $\Rey = 1000$, $\epsilon = 0.04$, $\delta = 0.1$, $\lambda = 0.1$ and $\gamma = 0.5$.}
\label{fig:NumericsVsAsymptoticsCompositeV}
\end{figure}
\begin{figure}
\centering
\vspace{0mm}
\begin{tabular}{ccc}
 & \quad \; (a) & \quad (b) \\[5mm]
    &  
    \begin{overpic}[width=0.4\textwidth,tics=10]
    {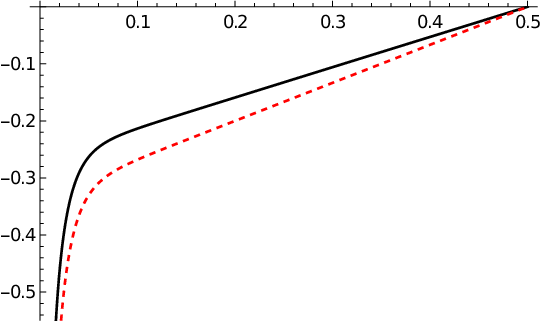}
    \put(51,62){$y$}
    \put(6,62){$v$}
    \put(3,22.8){\line(1,0){6}}
    \put(-17,21.4){$v^* = \mathlarger{\frac{1}{3}}$}
    \end{overpic} 
    &
    \begin{overpic}[width=0.42\textwidth,tics=10]
    {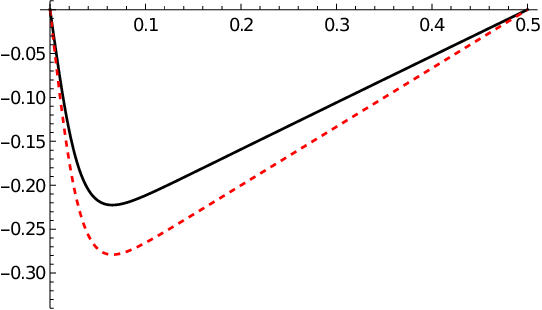}
    \put(52,59){$y$}
    \put(8,60){$v$}
    \put(5,1.5){\line(1,0){6}}
    \put(-17,0){$v^* = \mathlarger{\frac{1}{3}}$}
    \put(75,5){$\boxed{\epsilon = 0.1}$}
    \end{overpic}
    \\[2mm]
\end{tabular}
\vspace{2mm}
\caption{Comparison of the leading-order asymptotic composite solution~\eqref{eqn:vComposite} in red dashed lines, with the numerical solution in solid black lines, for the $y$-component of velocity, $v$. The solutions are taken at (a) $x = x_i$, the centre of the branched channel/point-sink and (b) $x = x_i + \epsilon/2$, the right-hand edge of the periodic unit cell. In this example, we consider the centremost cell, taking $x_i = 0.55$, and vary the cell width. Here, $\mathcal{P}_\text{out} = 0.4$, $\Rey = 1000$, $\epsilon = 0.1$, $\delta = 0.1$, $\lambda = 0.1$ and $\gamma = 0.5$.}
\label{fig:NumericsVsAsymptoticsCompositeV_eps01}
\end{figure}

\begin{figure}
    \centering
    \vspace{4mm}
    \begin{overpic}[width=0.46 \textwidth,tics=10]
    {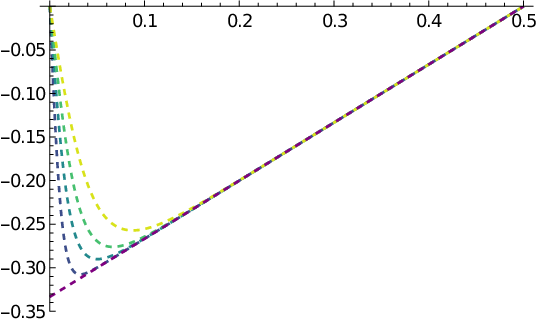}
    \put(53,62){$y$}
    \put(8,62){$v$}
    \put(22.3,30){\vector(-1,-2){17}}
    \put(-0.5,-8){$\epsilon \rightarrow 0$}
    \put(3.9,4.3){\line(1,0){5.5}}
    \put(-16,3){$v^* = \mathlarger{\frac{1}{3}}$}
    \end{overpic} 
    \vspace{5mm}
\caption{Leading-order asymptotic composite solution \eqref{eqn:vComposite} for $v$ at $x = x_i + \epsilon/2$ for $\epsilon = 0.04, \, 0.1, \, 0.125, \, 1/6$.  The outer solution for the velocity~(\ref{eqn:LOvOuter}) is indicated in purple. Here, $\mathcal{P}_\text{out} = 0.4$, $\Rey = 1000$, $\delta = 0.1$, $\lambda = 0.1$ and $\gamma = 0.5$. For varying $\epsilon$, we remain in the correct limit, $\epsilon \gg 1/\sqrt{\Rey}$.}
\label{fig:v_asymptotic solution}
\end{figure}
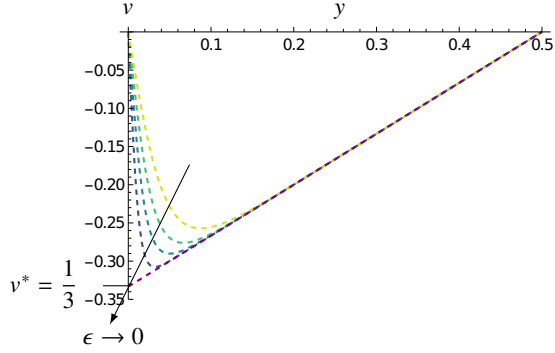

\begin{figure}
    \centering
    \vspace{3mm}
    \begin{tabular}{ccc}
    &  \quad \; (a) & \quad (b) \\[2mm]
    &  
    \begin{overpic}[width=0.44\textwidth,tics=10]
    {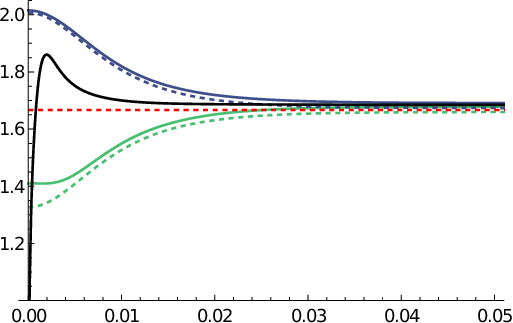}
    \put(50,-5){$y$}
    \put(4,67){$u$}
    \end{overpic}
    &
    \begin{overpic}[width=0.44\textwidth,tics=10]
    {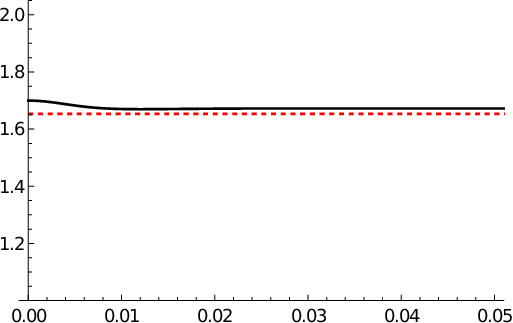}
    \put(50,-5){$y$}
    \put(4,67){$u$}
    \end{overpic}
\end{tabular}
    \vspace{5mm}
\caption{Comparison of the leading-order asymptotic composite solution~\eqref{eqn:compU} in dashed lines, with the numerical solution in solid lines, for the $x$-component of velocity, $u$. Here, we only plot close to $y=0$ as the solution is constant for larger $y$. The numerical solution in solid black lines and asymptotic solution in red dashed lines are taken at (a) $x = x_i$, the centre of the branched channel/point-sink and (b) $x = x_i + \epsilon/2$, the right-hand edge of the periodic unit cell. In this example, we consider the centremost cell, taking $x_i = 0.5$. The additional solutions in (a) are taken at $x=0.49$ (blue) and $x=0.51$ (green). Here, $\mathcal{P}_\text{out} = 0.4$, $\Rey = 1000$, $\epsilon = 0.04$, $\delta = 0.1$, $\lambda = 0.1$ and $\gamma = 0.5$.}
\label{fig:u495051}
\end{figure}

\begin{figure}
    \centering
    \vspace{6mm}
    \begin{overpic}[width=0.43\textwidth,tics=10]
    {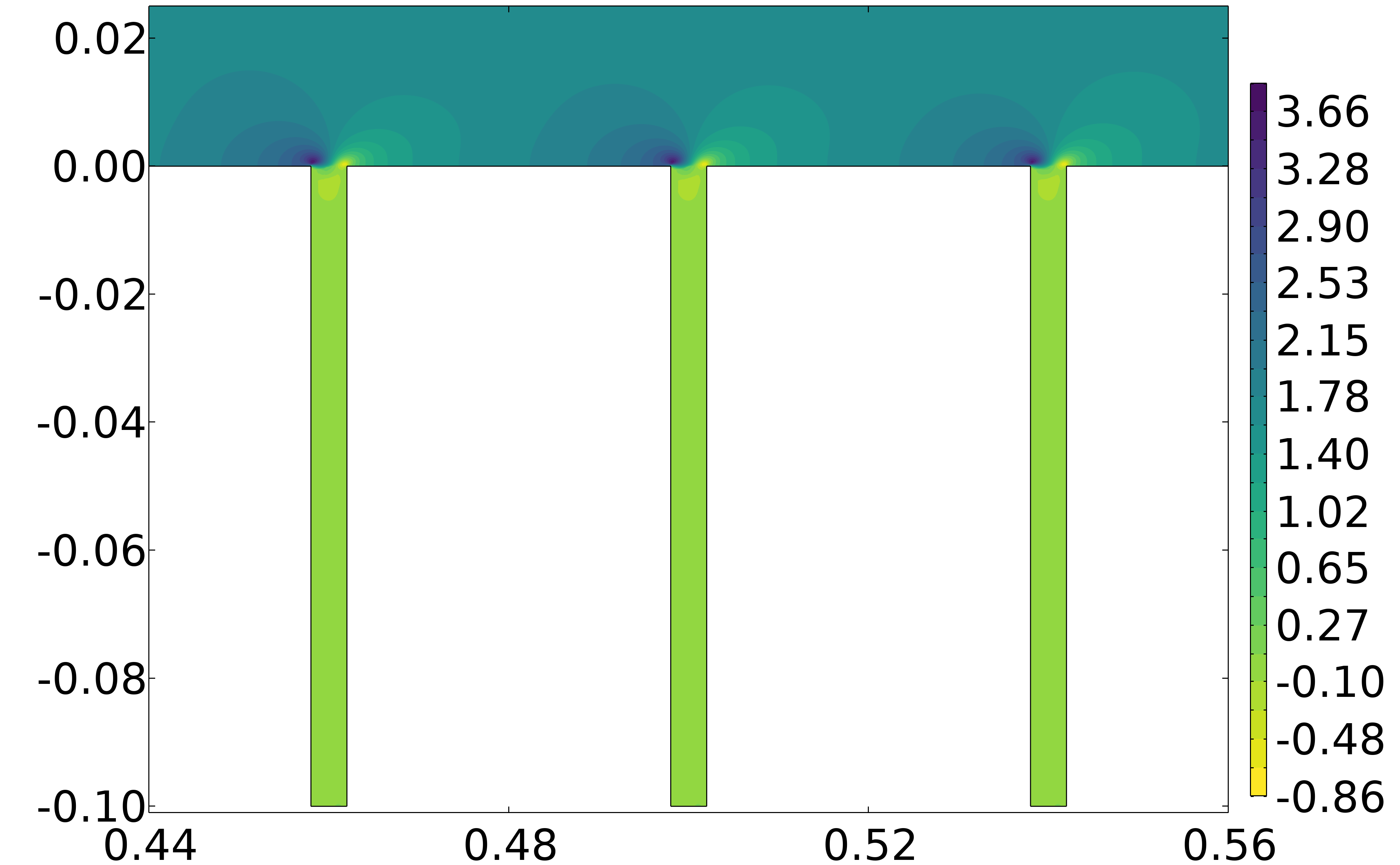}
    \put(47.75,-5.5){$x$}
    \put(-2.5,30){$y$}
    \put(102,30){$u$}
    \put(46.5,67){(a)}
    \end{overpic}
    \hspace{10mm}
    \begin{overpic}[width=0.43\textwidth,tics=10]
    {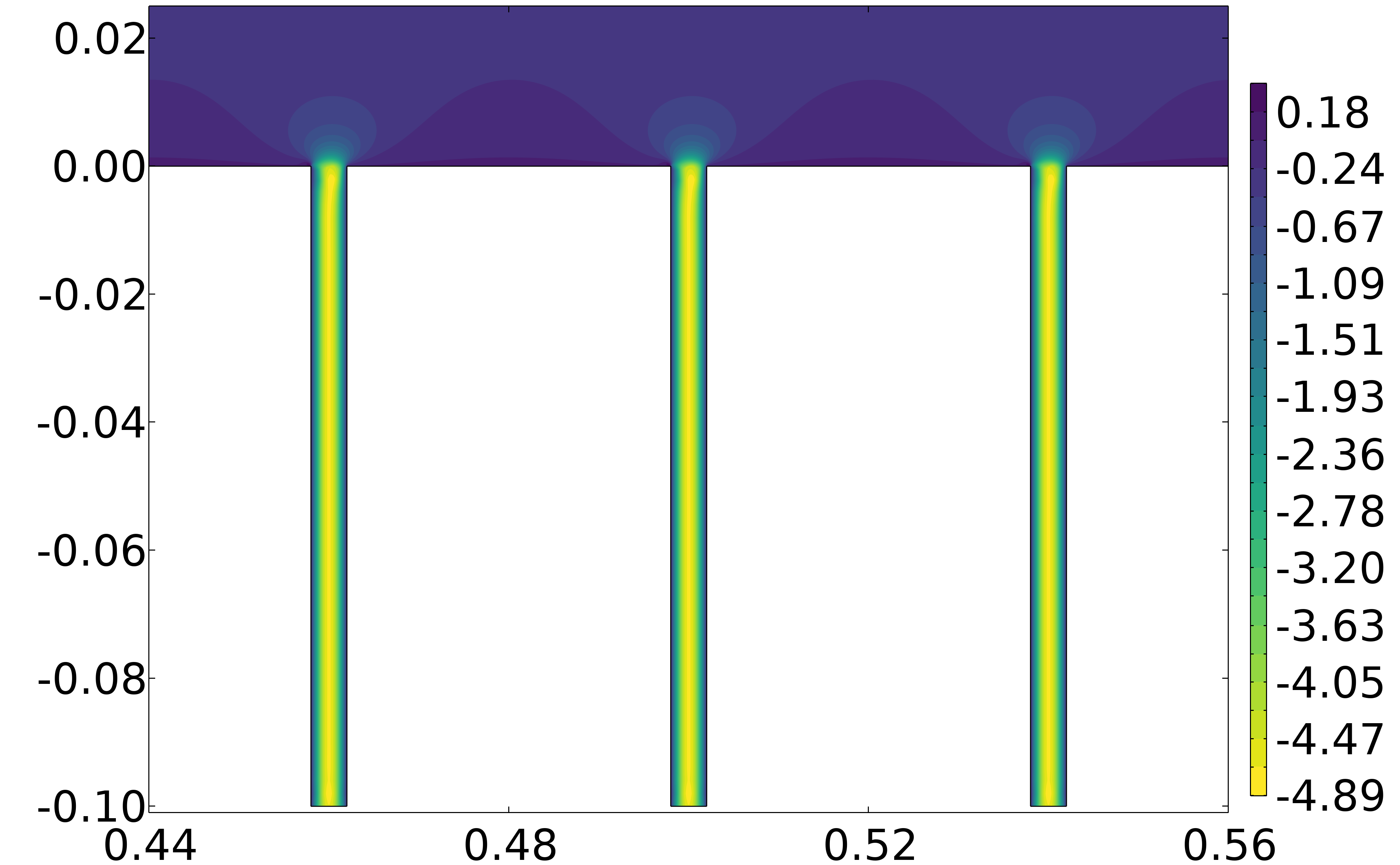}
    \put(47.75,-5.5){$x$}
    \put(-2.5,30){$y$}
    \put(102,30){$v$}
    \put(46.5,67){(b)}
    \end{overpic}
    \vspace{4mm}
\caption{Numerical solutions, zoomed \change{into}{in to} individual branched channels, for (a) the $x$-component of the flow speed, $u$, and (b) the $y$-component of the flow speed, $v$. Here, $\mathcal{P}_\text{out} = 0.4$, $\Rey = 1000$, $\epsilon = 0.04$, $\delta = 0.1$, $\lambda = 0.1$ and $\gamma = 0.5$.}
\label{fig:ZoomedFlow}
\end{figure}

We compare our asymptotic composite solution for $v(x^*, y)$, given by (\ref{eqn:vComposite}), with the numerical solution by plotting $v$ versus $y$ at the centre of a cell, $x^* = x_i$ (figure~\ref{fig:NumericsVsAsymptoticsCompositeV}(a)) and at the right-hand edge of the cell, $x_i + \epsilon/2$, (figure~\ref{fig:NumericsVsAsymptoticsCompositeV}(b)), for $\epsilon=0.04$. We see that the asymptotic and numerical results agree well and that the agreement improves as $\epsilon$ decreases whilst keeping $\Rey$ fixed, as expected (see figures~\ref{fig:NumericsVsAsymptoticsCompositeV} and \ref{fig:NumericsVsAsymptoticsCompositeV_eps01}). We clearly see the presence of the boundary layer at $y=0$, with the solution getting sharper as $\epsilon$ decreases as seen in figure~\ref{fig:v_asymptotic solution}, highlighting that the leading-order outer flow is a good approximation for all $y$, as $\epsilon \rightarrow 0$.

We similarly compare the asymptotic solution for $u(x^*, y)$ given by (\ref{eqn:compU}) to the numerical solution along the centre of a cell, at $x^* = x_i$ (figure~\ref{fig:u495051}(a)), and at the right-hand edge of the cell, $x_i + \epsilon/2$, (figure~\ref{fig:u495051}(b)), for $\epsilon=0.04$. Once again, we see that the asymptotic and numerical results agree well\add{,} apart from in a region close to $y=0$. We notice a significant disagreement between the numerics and the asymptotics along the centre of a cell. In figure~\ref{fig:u495051}(a), we plot additional $u(x^*, y)$ for $x^*$ between the edge of the branched channel and the edge of the cell. For these additional values, the asymptotics agree reasonably well with the numerics which suggests the disagreements are due to the finite size of the channel entrance. This error would reduce by decreasing the width of the channels in the numerical simulations. We note in passing, in figure~\ref{fig:ZoomedFlow}(a), that $u$ drops below zero around the right-hand corner of the branched channel entrance indicating a region of flow reversal.

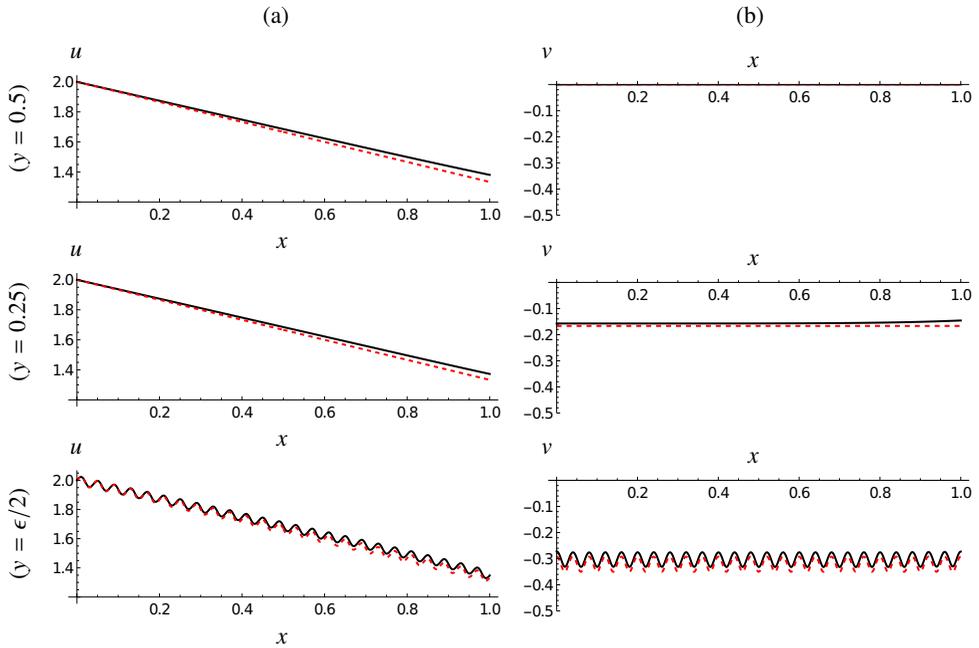
\begin{figure}
\centering
\vspace{2mm}
\begin{tabular}{ccc}
 & \,(a) & \, (b) \\[6mm]
    \begin{turn}{90}
        \qquad ($y = 0.5$)
        \vspace{3mm}
    \end{turn}
    &
    \begin{overpic}[width=0.44\textwidth,tics=10]
    {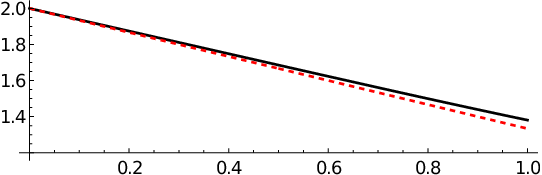}
    \put(4,36){$u$}
    \put(50,-6){$x$}
    \end{overpic} 
    &   
    \begin{overpic}[width=0.44\textwidth,tics=10]
    {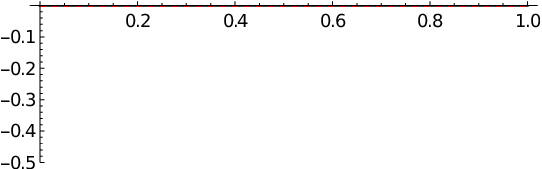}
    \put(4,36){$v$}
    \put(50,34){$x$}
    \end{overpic} 
    \\[6mm]
    \begin{turn}{90}
        \quad \; ($y = 0.25$)
        \vspace{3mm}
    \end{turn}
    &
    \begin{overpic}[width=0.44\textwidth,tics=10]
    {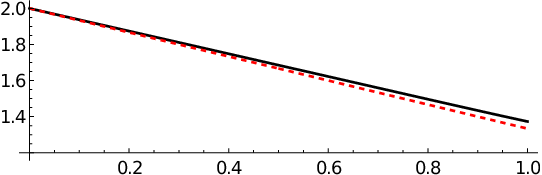}    
    \put(4,36){$u$}
    \put(50,-6){$x$}
    \end{overpic} 
    &   
    \begin{overpic}[width=0.44\textwidth,tics=10]
    {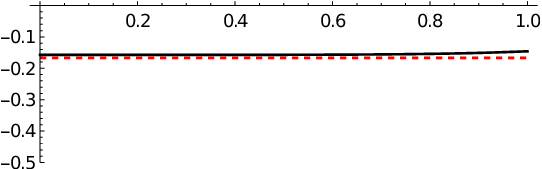}   
    \put(4,36){$v$}
    \put(50,34){$x$}
    \end{overpic} 
    \\[6mm]
    \begin{turn}{90}
        \quad \; ($y = \epsilon/2$)
        \vspace{3mm}
    \end{turn}
    &
    \begin{overpic}[width=0.44\textwidth,tics=10]
    {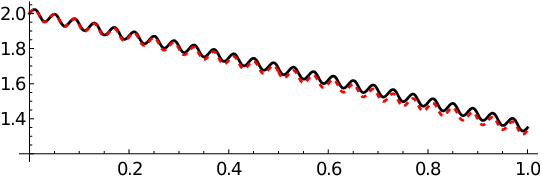}       
    \put(4,36){$u$}
    \put(50,-6){$x$}
    \end{overpic} 
    &   
    \begin{overpic}[width=0.44\textwidth,tics=10]
    {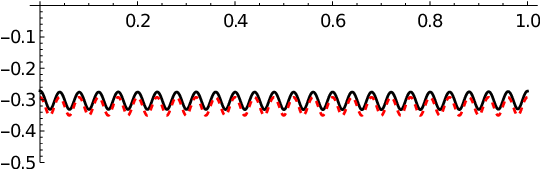}    
    \put(4,36){$v$}
    \put(50,34){$x$}
    \end{overpic} 
\end{tabular}
\vspace{5mm}
\caption{Comparison of the leading-order asymptotic composite solution in red dashed lines, with the numerical solution in solid black lines, for (a) the $x$-component of velocity, $u$, given by (\ref{eqn:compU}) and (b) the $y$-component of velocity, $v$, given by (\ref{eqn:vComposite}). The solutions are taken at $y = \epsilon/2$, $0.25$, and $\gamma$. Here, $\mathcal{P}_\text{out} = 0.4$, $\Rey = 1000$, $\epsilon = 0.04$, $\delta = 0.1$, $\lambda = 0.1$ and $\gamma = 0.5$.}
\label{fig:NumericsVsAsymptotics_u_composite}
\end{figure}

In figure~\ref{fig:NumericsVsAsymptotics_u_composite}(a), we compare the asymptotic solution for $u(x,y^*)$ given by (\ref{eqn:compU}) with the numerical solution, for three values of $y^*$. We observe excellent agreement between the asymptotic and numerical solutions. We further see minimal $y$-dependence in both the numerical results and composite solution away from $y=0$. As we approach the wall, we see that, in the boundary layer, the asymptotic solution accurately captures the oscillations seen in the numerics. Our results indicate that the discrete behaviour of the branched channels is confined to the inner region and the outer flow only sees this as an effective boundary condition. We similarly compare the asymptotic and numerical solutions for $v(x, y^*)$ given by (\ref{eqn:vComposite}) in figure~\ref{fig:NumericsVsAsymptotics_u_composite}, and again see excellent agreement for the three values of $y^*$.

\begin{figure}
    \centering
    \vspace{10mm}
    \hspace{-4mm}
    \begin{overpic}[width=0.4\textwidth,tics=10]
    {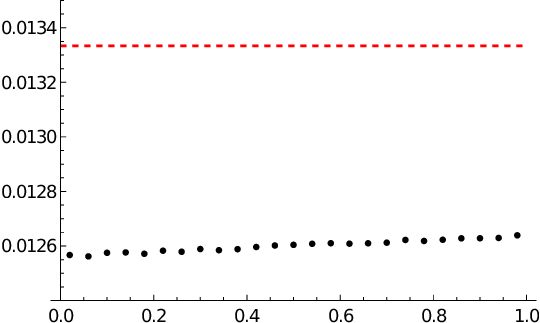}
    \put(51,73){(a)}
    \put(9,64){$Q^\text{branch}_i$}
    \put(53,-6){$x$}
    \put(0,51.35){\color{red}\line(1,0){11.5}}
    \put(-17.5,51){\textcolor{red}{$\epsilon \kappa \mathcal{P}_\text{out}$}}
    \end{overpic} 
    \hspace{2mm}
    \begin{overpic}[width=0.4\textwidth,tics=10]
    {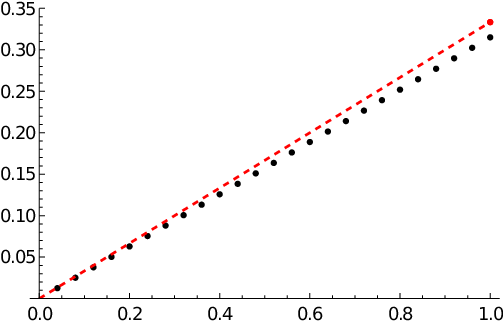}
    \put(51,73){(b)}
    \put(51,-6){$x$}
    \put(6,67){$Q^\text{cumulative}$}
    \put(76.5,62){\textcolor{red}{$(1,1/3)$}}
    \put(97.5,49){$(1,0.315)$}
    \end{overpic} 
    \vspace{3mm}
\caption{Comparison of the (a) flux through each branched channel, recorded at $x = x_i$, and (b) cumulative flux through each branched channel, recorded at $x = \epsilon i$, for $i = 1, 2, \cdots, N$. We plot the numerical solution of the flux (black dots), the asymptotic flux (red dashed lines) given by (\ref{eqn:DimlessFluxStrength}) and (\ref{eqn:IntegralFlux}), and the flux (\ref{eqn:DimlessFluxStrength}) using the pressure found numerically at the top of each branched channel, $p(x_i, 0)$, (blue dots). Here, $\mathcal{P}_\text{out} = 0.4$, $\Rey = 1000$, $\epsilon = 0.04$, $\delta = 0.1$, $\lambda = 0.1$ and $\gamma = 0.5$.}
\label{fig:EachHoleComp}
\end{figure}
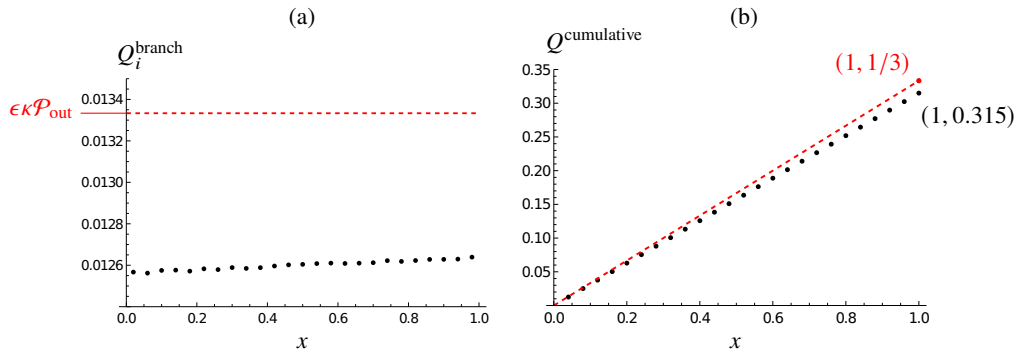

In figure~\ref{fig:EachHoleComp}(a), we compare the flux through each branched channel found numerically with the leading-order flux found in (\ref{eqn:DimlessFluxStrength}), and observe an $\mathcal{O}(\epsilon^2)$ discrepancy. Integrating the effective boundary condition~(\ref{eqn:WallLOFlux}) over $x$, we find that the cumulative flux along the wall is given by
\begin{equation}
    Q^\text{cumulative}(x) = \kappa \mathcal{P}_\text{out} x.
    \label{eqn:IntegralFlux}
\end{equation}
In figure~\ref{fig:EachHoleComp}(b), we compare this with the cumulative flux through each branch found numerically, and again we see excellent agreement, albeit with an error that grows linearly since we are integrating values that have systematic and asymptotically small errors. 

\subsection{Solution effects due to angled branches}

\label{section:AngledChannels}

\begin{figure}
\centering
\vspace{1mm}
\begin{overpic}[width=0.9\textwidth]{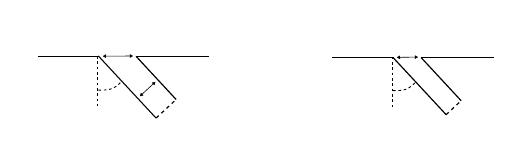}
    \put(20,24){(a)}
    \put(28.25,9){$\delta \epsilon$}
    \put(21,9){$\alpha$}
    \put(21.5,18.5){$w$}
    \put(75,24){(b)}
    \put(75.5,18.5){$\delta \epsilon$}
    \put(76.25,8.75){$\alpha$}
    \end{overpic}
    \vspace{-5mm}
    \caption{Angled geometry with (a) constant branch width $\delta \epsilon$ and hole width $w$, and (b) constant hole width $\delta \epsilon$. The angle, $\alpha$, is defined between the branched channels walls and the negative $y$-axis.}
    \label{fig:AngledStrucutres}
\end{figure}
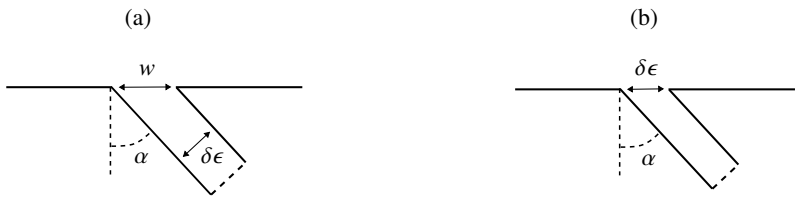

We now consider the effect of angling the branched channels. We denote the angle between the branched channel walls and the negative $y$-axis by $\alpha$, with $\alpha > 0$ corresponding with channels that are angled in the overall flow direction. We consider the cases where either (a) the branched channel width stays constant as $\alpha$ increases, or (b) the hole size stays constant as $\alpha$ increases, as shown in figure~\ref{fig:AngledStrucutres}. 

Before we consider each case, as in \S\ref{section:BranchedFlow}, we draw attention to the behaviour around the entrance to the branched channels. We have assumed that the flow within the branched channels is unidirectional, but, at the entrance to the branched channel, there is a small region where the flow will not be fully developed. Once again, we neglect any complex behaviour in this region (see Appendix~\ref{appA}). Under this assumption, and given that the resistance to the flow is {\it along} the channel, varying $\alpha$ while keeping the branch width $\delta\epsilon$ constant does not affect the boundary condition~(\ref{eqn:OuterEffective}). Therefore the effective flux~(\ref{eqn:WallLOFlux}) stays the same. The only requirement is that the hole size, $w$, is sufficiently small when compared to the cell width, such that we may approximate the hole by a point-sink. This means that we require
\begin{equation}
    w \ll \epsilon \quad \Rightarrow \quad \delta \ll \cos{\alpha}.
    \label{eqn:alphaRestriction}
\end{equation}

On the other hand, holding the hole size constant and varying $\alpha$ changes the width of the branched channel to $\delta \epsilon \cos{\alpha}$, resulting in
\begin{equation}
    v^* = - \frac{\delta^3 \Rey \cos^3{\alpha}}{12 \lambda} \mathcal{P}_\text{out},
\end{equation}
and so the total flux through the branched channels is
\begin{equation}
    Q = \frac{\delta^3 \Rey \cos^3{\alpha}}{12 \lambda} \mathcal{P}_\text{out}, \label{eqn:FluxConstHole}
\end{equation}
\textit{i.e.,} the flux decreases as $\alpha$ increases.

\begin{figure}
\vspace{2mm}
\centering
\begin{tabular}{cc}
\quad \quad (a) & \quad \quad (b) \\[6mm]
    \begin{overpic}[width=0.43\textwidth,tics=10]
    {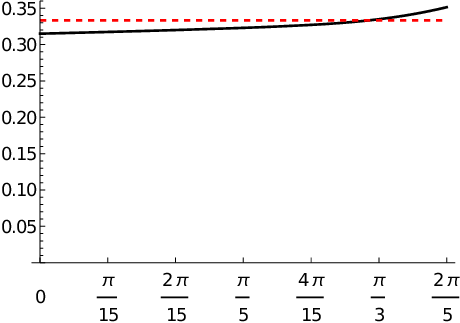}
    \put(52,-7){$\alpha$}
    \put(7,75){$Q$}
    \end{overpic} 
    \hspace{5mm}
    &   
    \begin{overpic}[width=0.43\textwidth,tics=10]
    {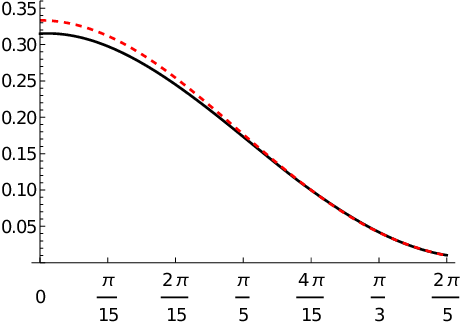}
    \put(52,-7){$\alpha$}
    \put(7,75){$Q$}
    \end{overpic}
\end{tabular}
\vspace{3mm}
\caption{Comparison of the total flux through the discrete branches, $Q$, found numerically (black solid) to the asymptotic prediction of the effective boundary condition (red dashed), for (a) constant branch channel width and (b) constant branch hole width. The asymptotic flux is given by (\ref{eqn:WallLOFlux}) and (\ref{eqn:FluxConstHole})\add{,} respectively. Numerical solutions are found similarly to figure~\ref{fig:FullModel_Flow}, with the same parameter values, but with the addition of the respective angled branched channels. Here, $\mathcal{P}_\text{out} = 0.4$, $\Rey = 1000$, $\epsilon = 0.04$, $\delta = 0.1$, $\lambda = 0.1$ and $\gamma = 0.5$.}
\label{fig:AngleComparison}
\end{figure}

We compare these predictions with their numerical equivalents as $\alpha$ is varied, in figure~\ref{fig:AngleComparison}. We restrict our attention to angles $\alpha \leq \pi /3$, for which $\delta = 0.1 \ll 0.5 < \cos{\alpha}$. The asymptotic solutions match the numerical simulations well in both cases, with an $\mathcal{O}(\epsilon)$ over-prediction, which decreases as $\alpha$ approaches $\pi/3$. As we increase $\alpha$ further, past $\pi/3$, we see an under-prediction in the constant width scenario, but continued agreement in the constant hole scenario. In both cases we are ultimately restricted by $\alpha < \pi/2$. A similar result to figure~\ref{fig:AngleComparison}(a) is found by \cite{LeakyChannels}, determining that the flux through the branched channel has minimal variation when changing the angle, $\alpha$, for a constant channel width.

\section{Particle model derivation}

\label{section:particle_model}

We now consider the behaviour of individual spherical particles moving through the device in the main channel, $(\hat{x}, \hat{y}) \in [0,L] \times [0, h_1]$. We suppose for simplicity that the only force acting on \change{the}{a} particle\delete{s} is fluid drag, \change{which we assume has quadratic form}{which has quadratic dependence on the velocity due to the high speed of the flow \citep{benchaita1983erosion}}. \delete{Writing $\mathbf{F} = m \boldsymbol{a}$ for a particle, w}We \change{have that}{follow \cite{VigoloTJunction} to establish a force balance for each particle, given by} $\change{\mathbf{F}}{\hat{\mathbf{F}}} = m \change{\boldsymbol{a}}{\hat{\boldsymbol{a}}}$ \add{, but significantly reduce the complexity by removing all forces other than drag. In doing this, we write}
\begin{equation}
    \frac{4}{3} \rho_p \pi a^3 \frac{\mathrm{d}^2 \hat{\boldsymbol{x}}_p}{\mathrm{d} \hat{t}^2} = \frac{C_D}{2} \rho_f \pi a^2 \, \bigg\rvert \, \hat{\boldsymbol{u}} - \dd{\hat{\boldsymbol{x}}_p}{\hat{t}} \bigg\rvert \left( \hat{\boldsymbol{u}} - \dd{\hat{\boldsymbol{x}}_p}{\hat{t}} \right),
    \label{eqn:DimalForce}
\end{equation}
where $\hat{\boldsymbol{x}}_p = (\hat{x}_p (\hat{t}), \hat{y}_p (\hat{t}))$ is the particle position, $a$ and $\rho_p$ denote the particle radius and density\add{,} respectively, $\hat{\boldsymbol{u}} = (\hat{u}, \hat{v})$ is the flow velocity, and $C_D$ is a drag coefficient (see, for example, \cite{benchaita1983erosion}). We release a particle at the inlet, $\hat{x}=0$, with position $\hat{y} = \hat{y}_0$ and assume that it has initial velocity given by the inlet flow velocity,
\begin{align}
    \hat{\boldsymbol{x}}_p \ownadd{(0)} &= (0, \hat{y}_0),     \label{eqn:particleInitialPos_Dimal}\\
    \dd{\hat{\boldsymbol{x}}_p}{\hat{t}} \delete{(0, \hat{y}_0)}\ownadd{(0)} &= \hat{\boldsymbol{u}} (0, \hat{y}_0).
    \label{eqn:particleInitialVel_Dimal}
\end{align}

We assume particle collisions with the walls are perfectly elastic, and model the ricochet effect via a simple bounce condition, \textit{i.e.,} angle of incidence $=$ angle of reflection, giving 
\begin{equation}
\left(\dd{\hat{x}_p}{\hat{t}}, \dd{\hat{y}_p}{\hat{t}} \right) \bigg \rvert^+ = \left(\dd{\hat{x}_p}{\hat{t}},  - \dd{\hat{y}_p}{\hat{t}} \right) \bigg \rvert^- 
\quad \text{on all walls,}
\label{eqn:ParticleBounceCondDimal}
\end{equation}
where $\cdot|^-$ means before the collision and $\cdot|^+$ means after the collision has occurred. 

\subsection{Non-dimensionalisation}

We nondimensionalise the model \eqref{eqn:DimalForce}--\eqref{eqn:ParticleBounceCondDimal} using the scalings
\begin{equation}
    \hat{\boldsymbol{x}}_p = L \boldsymbol{x}_p, \qquad \hat{\boldsymbol{u}} = \frac{\mathcal{Q}}{L} \boldsymbol{u}, \qquad \hat{t} = \frac{L^2}{\mathcal{Q}} t, \label{eqn:ParticleScalings}
\end{equation}
to find that
\begin{equation}
    \Sty \frac{\mathrm{d}^2 \boldsymbol{x}_p}{\mathrm{d} t^2} = \, \bigg\rvert \, \boldsymbol{u} - \dd{\boldsymbol{x}_p}{t} \bigg\rvert \left( \boldsymbol{u} - \dd{\boldsymbol{x}_p}{t} \right),
    \label{eqn:FMA_particle}
\end{equation}
where
\begin{equation}
    \Sty = \frac{8}{3} \frac{a \rho_p}{C_D L \rho_f}\delete{,}
\end{equation}
is the Stokes number. When $\Sty = 0$, particles are infinitely light tracer particles and follow the flow exactly. Otherwise, $\Sty > 0$ represents a particle with some non-zero density.

The inlet/initial conditions~\eqref{eqn:particleInitialPos_Dimal} and \eqref{eqn:particleInitialVel_Dimal} become
\begin{align}
    \boldsymbol{x}_p \ownadd{(0)} &= (0, y_0), \label{eqn:InitialParticlePos}\\
    \dd{\boldsymbol{x}_p}{t} \delete{(0, y_0)}\ownadd{(0)} &= \boldsymbol{u}(0, y_0), \label{eqn:InitialParticleVel}
\end{align}
where $y_0 = \hat{y}_0/L$, and the bounce condition~\eqref{eqn:ParticleBounceCondDimal} on the walls becomes
\begin{equation}
\left(\dd{{x}_p}{{t}}, \dd{{y}_p}{{t}} \right) \bigg \rvert^+ = \left(\dd{{x}_p}{{t}},  - \dd{{y}_p}{{t}} \right) \bigg \rvert^- .
\label{eqn:ParticleBounceCond}
\end{equation}

\subsection{Suitable flow field}

\label{section:particleSuitableField}

In the flow problem \add{in \S\ref{section:Model_Derivation}--\S\ref{section:CompSolution}}, we identified that there is an outer region away from the bottom wall, and an inner region of thickness $\epsilon$ close to the bottom wall. We found that the flow profile is given by the \delete{composition}\ownadd{composite} solution $\boldsymbol{u}_c = (u_c, v_c)$ shown in \eqref{eqn:compU} and \eqref{eqn:vComposite}. The effect of the wall only becomes noticeable at a substantial distance into the boundary layer, $y \approx \epsilon/2$. A key question is whether the outer flow provides a sufficient enough description of the flow to accurately capture the particle trajectories, or whether the (computationally more expensive) composite solution is required. To answer this question, we \delete{will} solve \eqref{eqn:FMA_particle}--\eqref{eqn:ParticleBounceCond} for several trajectories using (i) the outer flow, (ii) the composite flow, (iii) the full numerical flow and compare and contrast the results. We note that to ensure the flux out, $Q$, is the same in the asymptotic and numerical solutions, we must take $\mathcal{P}_\text{out} = 0.424$ in the numerical solution to compare to the asymptotic solution exactly, when $\mathcal{P}_\text{out} = 0.4$, so that $Q = 1/3$ in both. We explore two distinguished limits for the Stokes number, $\Sty = \mathcal{O}(1)$ and $\Sty=\mathcal{O}(\epsilon)$.

When simulating particle trajectories using the asymptotic results, we exploit the fact that the $y$-component of the velocity is an odd function in $y$, i.e., $v_c (x, -y) = - v_c (x,y)$. We therefore solve the governing equation~\eqref{eqn:FMA_particle} with inlet conditions \eqref{eqn:InitialParticlePos} and \eqref{eqn:InitialParticleVel} in the domain $[0,1] \times [-\gamma, \gamma]$ allowing the particle to pass through $y=0$ unless the trajectory enters a branched channel. When a particle passes through the wall at $y=0$, the angle of incidence is equal to the angle of reflection. We then take the modulus of the $y$-component of the trajectory, which ensures the bounce condition is satisfied for $y \geq 0$. As noted, this would be unsuitable if a particle trajectory enters a branched channel --- in our comparisons below, the specific values of $y_0$ are chosen to ensure that each particle lands away from a branched channel on every bounce.

\bigskip

\begin{figure}
\centering
\vspace{0mm}
\begin{tabular}{ccc}
    \begin{overpic}[width=0.4\textwidth,tics=10]
    {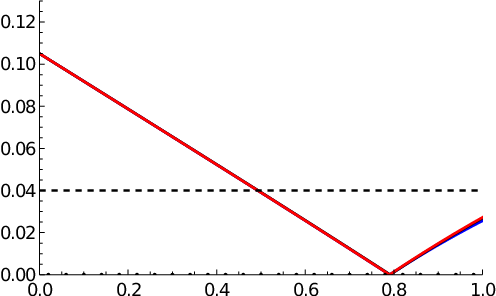}
    \put(65,50){$\boxed{y_0 = 0.105}$}
    \put(6.5,63){$y$}
    \put(51,-5){$x$}
    \end{overpic}
    &
    &
    \begin{overpic}[width=0.4\textwidth,tics=10]
    {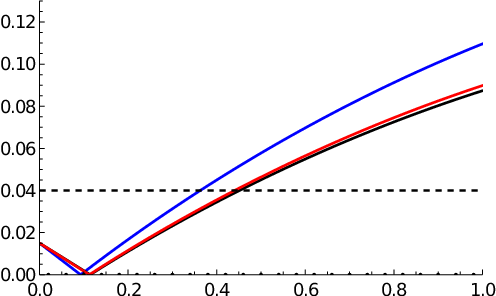}
    \put(15,50){$\boxed{y_0 = 0.015}$}
    \put(6.5,63){$y$}
    \put(51,-5){$x$}
    \end{overpic}    
\end{tabular}
\vspace{1mm}
\caption{Examples of possible bounces for $\Sty = 1$, with the flow field given by the outer solution $\boldsymbol{u}^{(o)}_0 (x,y)$ (blue), the composite solution $\boldsymbol{u}_c (x,y)$ (red) and the full numerical flow (black) from figure~\ref{fig:FullModel_Flow}. We vary the initial position, $y_0$ to show the minimal difference between trajectories when $\Sty = \mathcal{O}(1)$, for particles initialise both inside and outside of the boundary layer indicated by black dashed lines. The location of the entrances to the branched channels are indicated by the semicircles on $y=0$. Here, $\mathcal{P}_\text{out} = 0.4$ in the asymptotic solution and $\mathcal{P}_\text{out} = 0.424$ in the numerical simulation so that $Q = 1/3$ in both. We also have $\Rey = 1000$, $\epsilon = 0.04$, $\delta = 0.1$, $\lambda = 0.1$ and $\gamma = 0.5$.}
\label{fig:Stokes1}
\end{figure}
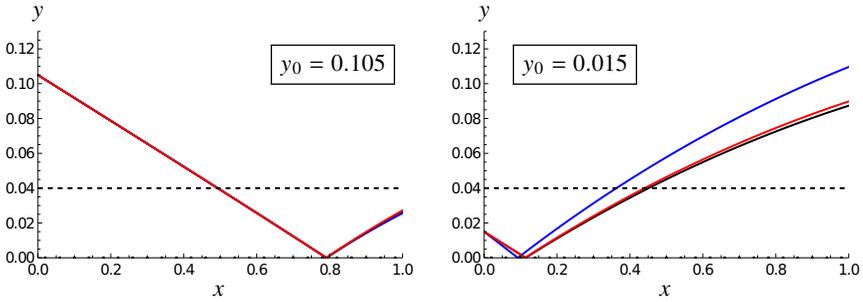

\noindent $\boldsymbol{\Sty = \mathcal{O}(1)}$:

\bigskip

\noindent We see in figure~\ref{fig:Stokes1}, in the limit when $\Sty = \mathcal{O}(1)$, there is minimal difference in the particle trajectories when using either flow. In both cases of $y_0$, the particle rapidly escapes the boundary layer before exiting out of the end of the main channel. Therefore, in this limit the boundary layer contribution has minimal effect on any particle which starts outside of the boundary layer and so using the outer flow is suitable in this limit. We note that when a particle starts within the boundary layer, the bounce trajectory does not compare so well, and so the outer solution is not so suitable. We also see similar behaviour when we introduce particles into the full branched channel flow, indicated in black in figure~\ref{fig:Stokes1} --- they too rapidly escape the boundary layer and exit out of the end of the main channel, following a similar path to the particles in the composite solution.

\bigskip

\begin{figure}
\centering
\vspace{0mm}
\begin{tabular}{ccc}
    \begin{overpic}[width=0.4\textwidth,tics=10]
    {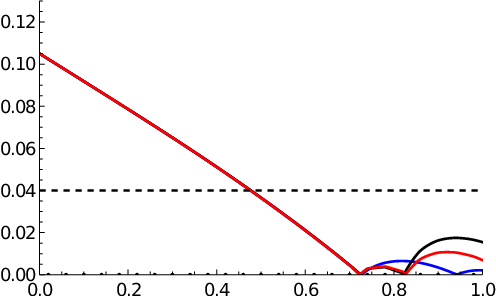}
    \put(65,50){$\boxed{y_0 = 0.105}$}
    \put(6.5,63){$y$}
    \put(51,-5){$x$}
    \end{overpic}
    &
    &
    \begin{overpic}[width=0.4\textwidth,tics=10]
    {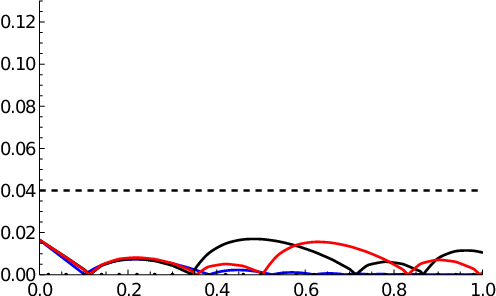}
    \put(15,50){$\boxed{y_0 = 0.0165}$}
    \put(6.5,63){$y$}
    \put(51,-5){$x$}
    \end{overpic}
\end{tabular}
\vspace{1mm}
\caption{Examples of possible bounces for $\Sty = 0.04$, with the flow field given by the outer solution $\boldsymbol{u}^{(o)}_0 (x,y)$ (blue), the composite solution $\boldsymbol{u}_c (x,y)$ (red) and the full numerical flow (black) from figure~\ref{fig:FullModel_Flow}. We vary the initial position, $y_0$ to show the minimal difference between trajectories when $\Sty = \mathcal{O}(\epsilon)$, for particles initialise both inside and outside of the boundary layer indicated by black dashed lines. The location of the entrances to the branched channels are indicated by the semicircles on $y=0$. Here, $\mathcal{P}_\text{out} = 0.4$ in the asymptotic solution and $\mathcal{P}_\text{out} = 0.424$ in the numerical simulation so that $Q = 1/3$ in both. We also have $\Rey = 1000$, $\epsilon = 0.04$, $\delta = 0.1$, $\lambda = 0.1$ and $\gamma = 0.5$.}
\label{fig:StokesEps}
\end{figure}
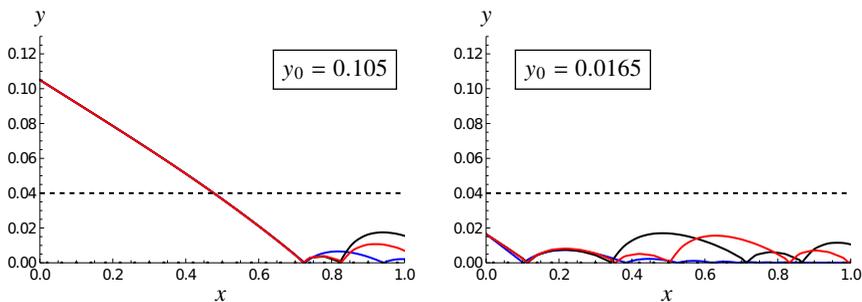

\noindent $\boldsymbol{\Sty = \mathcal{O}(\epsilon)}$:

\bigskip

\noindent We see in figure~\ref{fig:StokesEps} that the particle trajectories calculated using the outer flow, the composite flow, and the numerical solution no longer follow the same path (although they do initially), and there are significantly more bounces than in the $\Sty=\mathcal{O}(1)$ case. We see that, once a particle enters into the $\epsilon$-boundary layer, the Stokes number is small enough such that the particle motion is dominated by the flow and the particle remains within the boundary layer. Thus, in this limit, the composite solution should be used rather than the simpler outer flow. However, the agreement between these two flow fields and also the numerical solution will improve as $\epsilon \rightarrow 0$, as expected. Finally, in figure~\ref{fig:StokesEps}, we also observe that the particle trajectories in the composite flow seem disjoint or unresolved around the point-sinks, however this behaviour is confirmed by similar behaviour in the full numerical solution of the flow and particles.

In the results in section \ref{section:particle_results}, due to the previous observations, we solve for the particle trajectories using the composite flow \eqref{eqn:compU} and \eqref{eqn:vComposite} only, to best mimic particles in the full branched channel flow for all values of $\Sty$.

\section{Particle results}

\label{section:particle_results}

In branched channel filters, we are interested in diverting flow through the branched channels whilst retaining particles in the main channel flow. Since we have found an asymptotic prediction for the fluid flow through the branched channels, we now seek a similar result for particles. To quantify this, we define
\begin{equation}
    \mathcal{K} = \frac{\text{Number of particles left through the branched channels}}{\text{Number of particles at the inlet}}.
    \label{eqn:GeneralK}
\end{equation}

Ignoring particle--particle interactions, we release $i=19,999$ particles at slightly perturbed initial points $(0, y_0^j)$ around $i$ equispaced internal points (with step size $1/20,000$), for $j = 1, ..., i$ at $t=0$. The simulation terminates when all of the particles have left the system. We run 60 such simulations and average the value of $\mathcal{K}$ over them (noting that averaging over completely random $y_0$ positions achieves a similar result). We assume that the particles may bounce along the bottom wall, $y=0$, until they either leave through the end of the device or down a branched channel. Given that we have taken the limit $\delta \rightarrow 0$ in the flow problem, we need to impose a ``removal condition" that takes account of the finite size of the branched channels (otherwise we will significantly underestimate the number of particles that exit through these channels). We assume that, if a particle collides with the wall within $\delta \epsilon / 2$ either side of a point-sink, we remove the particle from the simulation and assume that the particle will have gone down a branched channel (see figure~\ref{fig:Particle_Point_Sink} for an example particle removal via a branched channel). To calculate the proportion of particles, $\mathcal{K}$, we count the number of particles that go down a branched channel as described and divide by $i$.

\begin{figure}
\vspace{1mm}
\centering
    \begin{overpic}[width=0.5\textwidth,tics=10]{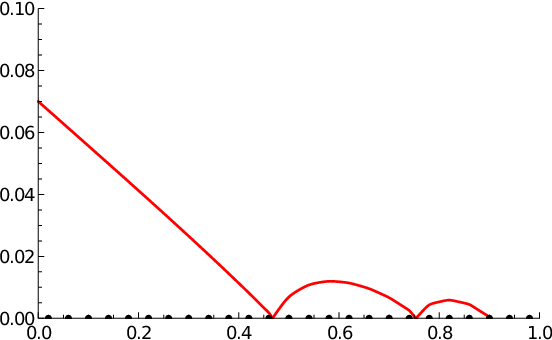}
    \put(6.5,64){$y$}
    \put(51,-4){$x$}
    \end{overpic}
\vspace{1mm}
\caption{Example trajectory in the composite solution for the flow \eqref{eqn:compU} and \eqref{eqn:vComposite}, when hitting a point-sink, or within a $\delta \epsilon / 2$ radius on $y=0$. The point-sinks are indicated by point markers on $y=0$. Here, $\epsilon = 0.04$, $\delta = 0.1$, $\gamma = 0.5$, $Q = 1/3$ and $y_0 = 0.07$.}
\label{fig:Particle_Point_Sink}
\end{figure}
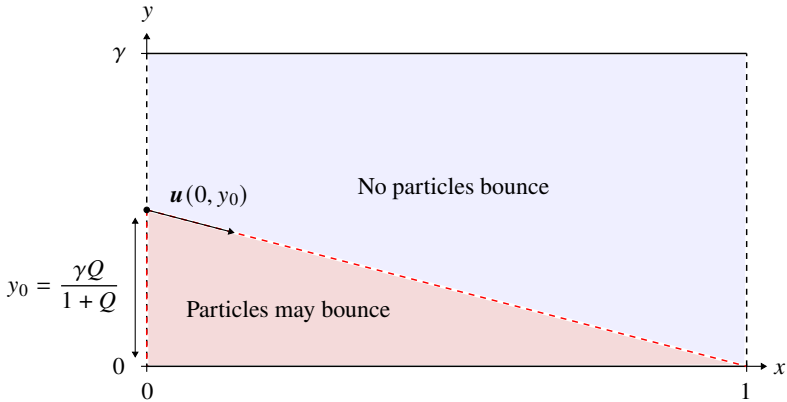
\begin{figure}
\vspace{1mm}
\centering
    \begin{overpic}[width=0.8\textwidth,tics=10]{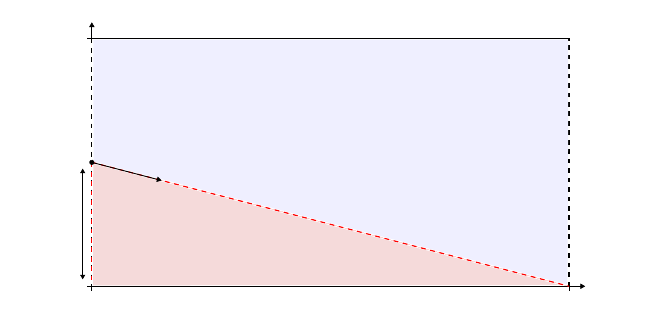}
    \put(-3,15){$y_0 = \mathlarger{\frac{\gamma Q}{1 + Q}}$}
    \put(91,5){$x$}
    \put(13.5,2){$0$}
    \put(86.75,2){$1$}
    \put(13.5,48.5){$y$}
    \put(10,43.5){$\gamma$}
    \put(10,5){$0$}
    \put(40,27){No particles bounce}
    \put(19,12){Particles may bounce}
    \put(17,26){$\boldsymbol{u}(0, y_0)$}
    \end{overpic}
\vspace{-1mm}
\caption{Phase diagram for $\Sty = \infty$, with the limiting trajectory (red dashed) having inlet position $\boldsymbol{x}_p = (0, y_0)$ and velocity $\boldsymbol{u}(0, y_0)$, for whether a particle will bounce.}
\label{fig:StInfinity}
\end{figure}

We start by calculating $\mathcal{K}$ for two distinct limits, $\Sty = 0$ and $\Sty = \infty$. These correspond, respectively, to infinitely light tracer particles and infinitely heavy particles that travel ballistically. 

When $\Sty = 0$, particles follow the streamlines exactly, exiting the main channel either down one of the branched channels or out of the end of the main channel. No particle will deviate from the flow or bounce on the bottom wall. Hence, we have that $\mathcal{K} = Q$. 

\begin{figure}
\vspace{1mm}
\centering
    \begin{overpic}[width=0.5\textwidth,tics=10]{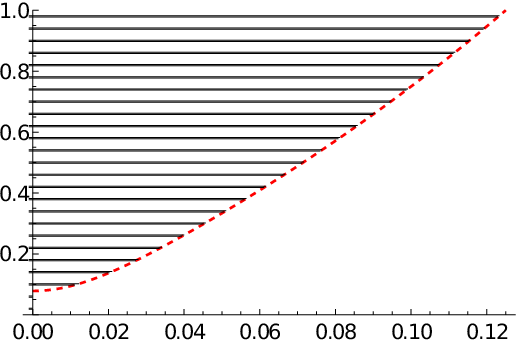}
    \put(50,-5){$y_0$}
    \put(0,68){$x$-intercept}
    \end{overpic}
\vspace{3mm}
\caption{The $x$-intercept, calculated using \eqref{eqn:xIntercept}, for a ballistic particle given an inlet position $y_0$, in the case where $\Sty=\infty$. We show the position of the branched channels by black horizontal lines, extended along to the $y_0$ values for which the $x$-intercept falls at the centre of the branched channel. Note that $y_0 \ne 0$ and so the first $x$-intercept is after the first two branched channels.}
\label{fig:Particle_Intercept}
\end{figure}

When $\Sty = \infty$, particles retain their initial velocity, and so follow the trajectory given by
\begin{equation}y~= ~y_0 + \frac{v_c(0, y_0)}{u_c (0, y_0)} x.\label{eqn:CJWB1}
\end{equation}
Since the outer flow is a good approximation for the flow outside the $\epsilon$-boundary layer, the \emph{limiting trajectory} for a particle, which we define as the dividing line between trajectories for particles that exit through the main channel without bouncing, and those that either exit through the branches or bounce then exit through the end, has the inlet position $y_0 = \gamma Q / 1+Q$ (see figure \ref{fig:StInfinity}). Particles released above this point will never bounce for any $\Sty$, indicating that a large proportion of particles will simply flow out of the main channel outlet. One might naively think that, since the ratio of total hole size to wall length is $\delta$, a $\delta$ proportion of these ballistic particles will enter the branched channels. This would be true if particles collide with the bottom wall at every $x \in [0,1]$, but it slightly overestimates $\mathcal{K}$ at $\Sty = \infty$, since there are locations on the bottom wall close to the inlet which cannot be reached by any ballistic trajectory. We calculate from \eqref{eqn:compU}--\eqref{eqn:vComposite} that
\begin{equation}
    \frac{v_c(0, y_0)}{u_c (0, y_0)} = - \gamma Q \left( \tanh{ \left( \frac{\pi y_0}{\epsilon} \right)} - \frac{y_0}{\gamma} \right),
\end{equation}
which we use with \eqref{eqn:CJWB1} to find the intercept position on the bottom wall of a ballistic particle released from $y_0$ at the inlet, given by 
\begin{equation}
    x_\text{intercept} = \frac{y_0}{\gamma Q \left( \tanh{ \left( \frac{\pi y_0}{\epsilon} \right)} - \frac{y_0}{\gamma} \right)}.
    \label{eqn:xIntercept}
\end{equation}
We calculate the range of $y_0$ values which have an $x$-intercept value that corresponds with the location of the branched channel (see figure \ref{fig:Particle_Intercept}). We then divide this total amount by $\gamma$ to normalise with the inlet height to give us the limiting value of $\mathcal{K}$ when $\Sty = \infty$. For the parameter values chosen, this value is $0.024$ (to $3$ decimal places). We note that this value will change depending on the initial velocity of the particles, and is slightly smaller than $\mathcal{K}=\delta Q / (1 + Q)$ --- the value we would have found from using the outer flow. Thus, we see that for small $\Sty$, the filter allows the same fraction of particles out through the branches as fluid flow, while for large $\Sty$, the filter retains more particles in the main channel flow, as is preferable.

\begin{figure}
\vspace{0mm}
\centering
\begin{tabular}{cc}
(a) & (b)\\[4mm]
    \begin{overpic}[width=0.4\textwidth,tics=10]
    {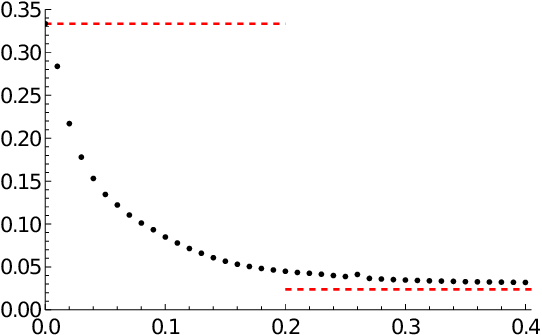}
    \put(50.25,-6){$\Sty$}
    \put(5.5,67){$\mathcal{K}$}
    \put(54.5,55.5){$\mathlarger{\frac{1}{3}}$}
    \end{overpic}
    &
    \begin{overpic}[width=0.4125\textwidth,tics=10]
    {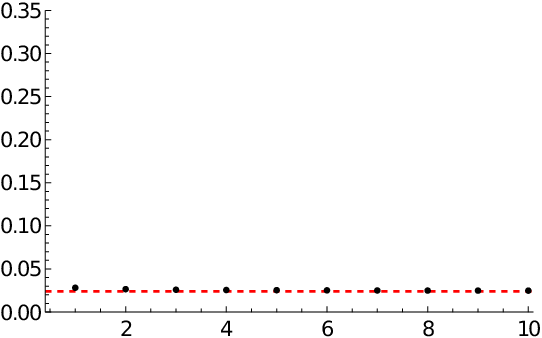}
    \put(50.25,-6){$\Sty$}
    \put(5.5,66){$\mathcal{K}$}
    \put(102,7){0.024}
    \end{overpic}
\end{tabular}
\vspace{1mm}
\caption{The proportion of particles, $\mathcal{K}$, leaving through the main channel compared to the total at the inlet, calculated using \eqref{eqn:GeneralK}, plotted as black points between (a) $\Sty \in [0, 0.4]$ with  a spacing of $0.01$ and (b) $\Sty \in [1,10]$ with a spacing of $1$. Here, we have taken $\epsilon = 0.04$, $\delta = 0.1$, $\gamma = 0.5$ and $Q = 1/3$. The limiting values for $\mathcal{K}$ are shown by red dashed lines. An explanation of the behaviour near $\Sty = 0.26$ is given in Appendix~\ref{appC}.}
\label{fig:KPlots}
\end{figure}

For a given $0 < \Sty < \infty$, we solve \eqref{eqn:FMA_particle}--\eqref{eqn:ParticleBounceCond} numerically using \texttt{Mathematica}, and calculate $\mathcal{K}$ from these simulations using
\eqref{eqn:GeneralK}. We show the dependence of $\mathcal{K}$ on $\Sty$ and the limiting behaviour when $\Sty = 0$ and $\Sty = \infty$ in figure~\ref{fig:KPlots}. We see that the number of particles that travel through the branched channels decreases as $\Sty$ increases, since more are deflected back into the main channel flow as we might expect. A similar conclusion is found by \cite{MantaRayRicochetSeparation}, where particle retention in the main channel flow increases for increasing particle size and density. We further see the large $\Sty$ approximation for $\mathcal{K}$ readily approximates the actual $\mathcal{K}$ for much smaller values of $\Sty$ than we might have anticipated.
The ideal regime is when the number of particles leaving through the branches is small compared to the flux --- this allows this filter concept to remove some `clean' water whilst retaining the majority of the particles flowing into the dead-end filter at the end of this simplified domain. However, the lowest fraction of particles that flow down the branched channels in this model is $0.024$. We note that the $\mathcal{K}$ behaviour is non-monotonic, most visibly near $\Sty=0.026$. These are not numerical artifacts, but rather a dynamical result of the system. We comment further on this behaviour in Appendix \ref{appC}, where at $\Sty = 0.26$ we see what we are calling a \emph{grazing} point, or a type of separatrix, in the solution. But for the purpose of this study, we focus on the overall trend in $\mathcal{K}$.

\begin{figure}
\centering
\begin{tabular}{cc}
(a) & (b)\\[4mm]
    \begin{overpic}[width=0.4\textwidth,tics=10]
    {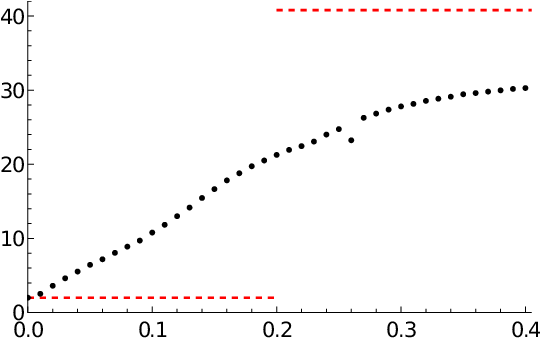}
    \put(50.25,-6){$\Sty$}
    \put(3,65){$\mathcal{R}$}
    \put(53,6.5){2}
    \end{overpic}
    &
    \begin{overpic}[width=0.4125\textwidth,tics=10]
    {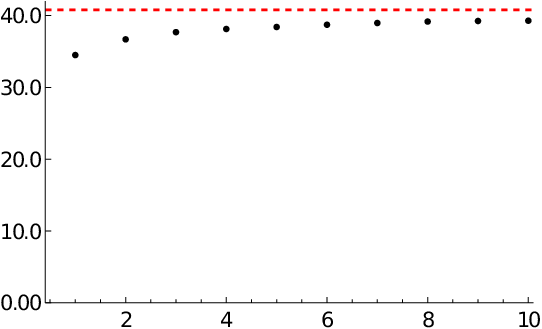}
    \put(50.25,-6){$\Sty$}
    \put(5,65){$\mathcal{R}$}
    \put(102,57){40.8}
    \end{overpic}
\end{tabular}
\vspace{1mm}
\caption{The proportion of particles, $\mathcal{R}$, leaving through the main channel rather than through the branched channels, calculated using \eqref{eqn:GeneralR}, and plotted as black points between (a) $\Sty \in [0, 0.4]$ with spacing $0.01$ and (b) $\Sty \in [1,10]$ with spacing $1$. We plot the corresponding values of $\mathcal{R}$ for figure~\ref{fig:KPlots}. Here, we have taken $\epsilon = 0.04$, $\delta = 0.1$, $\gamma = 0.5$ and $Q = 1/3$. The limiting values for $\mathcal{R}$ are shown by red dashed lines. An explanation of the behaviour near $\Sty = 0.26$ is given in Appendix~\ref{appC}.}
\label{fig:RPlots}
\end{figure}
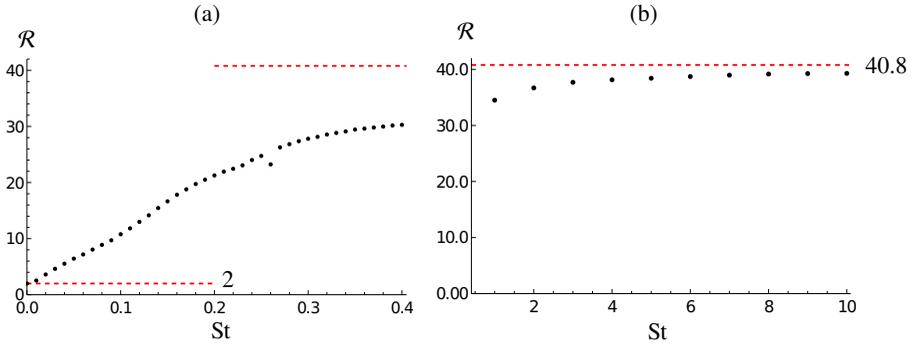

\begin{figure}
\centering
    \begin{overpic}[width=0.5\textwidth,tics=10]{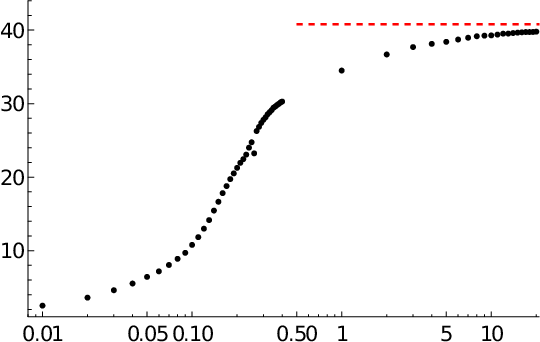}
    \put(53,-6){$\Sty$}
    \put(3.25,66){$\mathcal{R}$}
    \put(103,57.25){40.8}
    \end{overpic}
\vspace{3mm}
\caption{The proportion of particles, $\mathcal{R}$, leaving through the main channel rather than through the branched channels, calculated using \eqref{eqn:GeneralR}, and plotted as black points on a larger range of $\Sty$, on a $\log$ scale, over figure~\ref{fig:RPlots}. Here, we have taken $\epsilon = 0.04$, $\delta = 0.1$, $\gamma = 0.5$ and $Q = 1/3$. The limiting value at $\Sty = \infty$ for $\mathcal{R}$ is shown by a red dashed line.}
\label{fig:RPlotInf}
\end{figure}

Finally, we define
\begin{equation}
    \mathcal{R}=\frac{\text{fraction of particles left through the main channel}}{\text{fraction of particles left through the branched channels}} = \frac{1 - \mathcal{K}}{\mathcal{K}}.
    \label{eqn:GeneralR}
\end{equation}
We calculate
\begin{align}
    \Sty = 0 &: \quad \mathcal{R} = 2, \label{eqn:R1LimitSt}\\
    \Sty = \infty &: \quad \mathcal{R} \approx 40.8. \label{eqn:R2LimitSt}
\end{align}
We show how $\mathcal{R}$ varies with $\Sty$ in figure~\ref{fig:RPlots}. We see that $\Sty$ varies from $2$ to $40.8$ (for $Q=1/3$), showing that it is possible to divert a reasonable fraction of the flow through the branched channels whilst retaining a very high proportion of the particles exiting through the end of the main channel, even when $\Sty$ is relatively small. We see that it takes longer for $\mathcal{R}$ to converge to the large $\Sty$ asymptote than it does for $\mathcal{K}$; this is highlighted further in figure~\ref{fig:RPlotInf}.
Again, we highlight the unusual behaviour near $\Sty = 0.26$, and possibly other seemingly non-smooth points.

\section{Conclusions and extensions}

\label{section:conc}

In this paper, we consider a high-Reynolds-number flow of a viscous liquid in a branched channel filter, comprising of a series of T-junctions of width $\epsilon$. We prescribe a constant inlet flux, atmospheric pressure at the bottom of each branched channel, and \add{the} pressure at the main channel outlet. In addition to the existence of a viscous boundary layer, of thickness $1/\sqrt{\Rey}$, on the top and bottom wall of the main channel compartment, there is a layer of thickness $\epsilon$ on the bottom wall where the flow adjusts to go down the branched channels.

To simplify the full problem, we first consider the flow in each separate branched channel, finding the flux, $Q_i^{\textrm{branch}}$, as a function of the pressure at the top of the channel and the geometric and flow parameters. We take the limit as the channel width tends to zero and approximate the branch flow as point-sinks with strength $2 Q_i^{\textrm{branch}}$. In the main channel flow, we consider the regime where the branch separation is much larger than the viscous boundary layer, so that $\epsilon \gg 1/\sqrt{\Rey}$, and for simplicity, neglect the viscous boundary layers completely. We then decompose the domain into two regions\change{;}{:} one close to the wall where the discrete nature of the point-sinks are apparent, and a region away from the wall, where the effects of individual sinks are smoothed out --- the flow is inviscid in both regions. We find that the leading-order flow away from the wall has constant pressure.

We solve for the flow in an inner region close to $y=0$ using multiscale asymptotics to capture the rapid oscillations caused by the point-sinks. We utilise complex potential flow theory to find an explicit solution for the velocity. We match the solution to the outer flow, deriving an effective boundary condition that smooths out the discrete behaviour due to the branched channels. The dimensionless effective boundary condition is $v = - \delta^3 \Rey \mathcal{P}_\text{out}/12 \lambda$ on $y=0$, which is dependent on various design and operating parameters, but notably not on the T-junction width, $\epsilon$. 
Using this effective boundary condition, we solve for the leading-order outer flow velocity, and thus build a leading-order composite solution for the flow velocity that holds throughout the main compartment.

We validate our asymptotic solution by comparing with suitable numerical simulations for a large, finite $\Rey$.
We emulate inviscid main channel flow using a wall slip condition. We find good agreement between the asymptotic solution and numerical solution for an example set of parameters which improves as $\epsilon$ decreases (\textit{i.e.,} as the number of channels increases). The $\epsilon-$boundary layer smooths out the discrete behaviour of the point-sinks as required, and the effective boundary condition provides the main flow with the same necessary information as the full branched channel set-up. The explicit solution removes the need for computationally expensive simulations and thus has the potential to massively speed up the design process and optimisation of this device.

We explore how changing the branch angle affects the solution, for a constant branch width and constant hole size. We find that, for a constant branched channel width, which causes the branch hole size to alter with the angle, $\alpha$, the effective boundary condition remains constant for any angle provided the hole size, 
$\delta \epsilon \sec(\alpha) \ll \epsilon$ --- a similar result is found numerically by \cite{LeakyChannels}. On the other hand, for a constant hole size, the width of the branch channel changes with angle, $\alpha$, and so the effective boundary condition depends on $\alpha$. When comparing our asymptotic prediction to the numerical solution, we once again see good agreement.

Having solved for the flow in the simplified main channel domain, we then solve for the trajectories of individual spherical particles placed in the flow, assuming a wall bounce condition and the Stokes number, $\Sty$. We explore two distinguished limits between $\Sty$ and $\epsilon$ and conclude that simulating the particles in the leading-order outer flow does not mimic the effect of the branched channels well when $\Sty = \mathcal{O}(\epsilon)$ --- we must therefore use the leading-order composite flow.  To mimic particles flowing down the branched channels, a particle is removed from the simulation when it hits the bottom wall, $y=0$, at the location of a point-sink or within a $\delta \epsilon / 2$ distance away. We suppose that particles can bounce along the bottom wall an infinite number of times until they leave via the main outlet or via the bottom wall removal condition.  

We show that for zero Stokes number, the fraction of particles which flow down the branched channels is the same as the flux through the wall, $Q$, since the particles follow the flow through the point-sinks, and are therefore removed. For larger Stokes number, this limiting number falls to 0.024, slightly below the naive prediction of $\delta Q/(1+Q)$ as the particles become ballistic, i.e., $\Sty = \infty$. We further numerically calculate the fraction of particles which we lose via the bottom wall for $0 < \Sty < \infty$. We find that given these particular rules for spherical particles in this branched channel filter concept, the fraction of particles that are removed through the branched channels is much less than the fluid flux through the wall via point-sinks for $\Sty > 0$. Hence for particles to achieve the best trade-off between maximum fluid flux and minimal particles through the branched channels, spherical particles require a larger $\Sty$, which could correspond to flocculations of microfibres --- a similar conclusion was found by \cite{MantaRayRicochetSeparation} for plankton particles.

There are various extensions to this work that should be considered. Firstly, it might be useful to study the thus-far neglected viscous boundary layers to see whether they have any influence on the outer flow, rather than neglecting them. \add{This would introduce a non-zero vorticity near the branched channel entrances, which may change the dynamics of a particle's ricochet trajectory through rotation.} Secondly, one may consider the case where the viscous boundary layer is much larger than the $\epsilon-$boundary layer, and see how the multiple-scales process may be applied to find a suitable effective boundary condition in this regime. This would be another stepping stone towards our overarching aim of establishing effective boundary conditions for all regimes of laminar high\add{-}Reynolds\add{-}number flows, in addition to the low\add{-}Reynolds\add{-}number regime solutions given by \cite{LeakyChannels} --- which would negate the need for any numerics for the flow problem. 

When deriving our effective boundary condition, we also assumed that we are given the outlet pressure, $\hat{\mathcal{P}}_\text{out}$, as an operating parameter of the system (prescribed here as a constant, but in reality a function of time found by coupling the device to the downstream dead-end filter). Building and solving a model for the dead-end filter which incorporates clogging and caking of the filter surface, similarly to \cite{kory2021optimising}, would provide further insight on this pressure in order to explore its influence on the flow in the ricochet device. Finally, further research might consider more realistic, rounded-corner, geometries for the entrances to the branched channels, as they are in a ricochet device.

There are also a range of extensions to our \add{intentionally} simple particle model: firstly, we have thus far only included a drag term in our force balance model and so we should investigate and incorporate other important forces causing this ricochet effect; we ought to identify the correct physics for the bouncing condition by introducing a coefficient of restitution. This might be by considering a smaller microscale structure and studying how the particles ricochet back into the flow before feeding this into our simplified domain; changing the inlet condition for the particles (considering the trajectory shape before entering this device); incorporating particle-particle interactions between these spherical particles; finally, and possibly the most important, we have been modelling spherical particles in this paper. Although these \add{spherical particles} do mimic flocculations of microfibres, individual fibres are ultimately rod-like \add{(and possibly flexible)} and so the behaviours and forces would change. \cite{li2024dynamics} consider rod-like particles colliding and pole-vaulting over various obstacles. The ricochet separation concept might have improved efficiency with this pole-vaulting effect.

\change{Nevertheless t}{T}he work done in this paper is a crucial step into understanding the operation of branched channel filters and their effective use in washing machines. \add{Ultimately, it would be valuable to compare our results with those from experiments. Our work} \delete{It} has the potential to play a valuable role in optimising Beko's filtration devices and thus in preventing further microplastic pollution from entering  our already heavily polluted oceans.

\backsection[Acknowledgements]{We would like to thank Daniel Booth, Jon Chapman, Mo Dalwadi and Doireann O'Kiely for helpful mathematical discussions, and Graham Anderson, Konstantinos Pantelidis, Simge Tarku\c{c} and Melih Toklu for insightful discussions on microfibre filtration at Beko PLC and Ar\c{c}elik R\&D.}

\backsection[Funding]{T.F. is grateful to Beko PLC and EPSRC, grant reference number EP/W524311/1, for funding.}

\backsection[Copyright]{For the purpose of open access, the author has applied a CC BY public copyright licence to any author accepted manuscript arising from this submission.} 

\backsection[Declaration of interests]{The authors report no conflict of interest.}

\backsection[Author ORCIDs]{

\noindent T. Fastnedge, https://orcid.org/0009-0004-7266-7485;

\noindent C. J. W. Breward, https://orcid.org/0000-0003-4568-5261;

\noindent I. M. Griffiths, https://orcid.org/0000-0001-6882-7977.}

\appendix

\section{}\label{appA}

\begin{figure}
\centering
\begin{tabular}{cc}
    $y=0$
    &
    $y=-0.002$
    \\[1mm]
    \begin{overpic}[width=0.3\textwidth,tics=10]
    {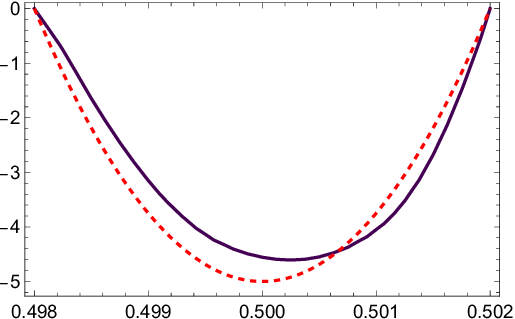}
    \put(-6,30){$v$}
    \end{overpic} 
    &   
    \begin{overpic}[width=0.3\textwidth,tics=10]
    {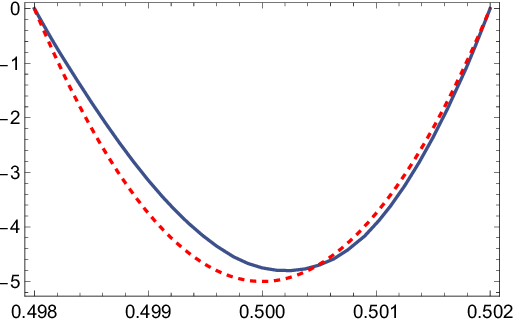}
    \end{overpic} 
    \\[5mm]
    $y=-0.004$
    &
    $y=-0.006$
    \\[1mm]
    \begin{overpic}[width=0.3\textwidth,tics=10]
    {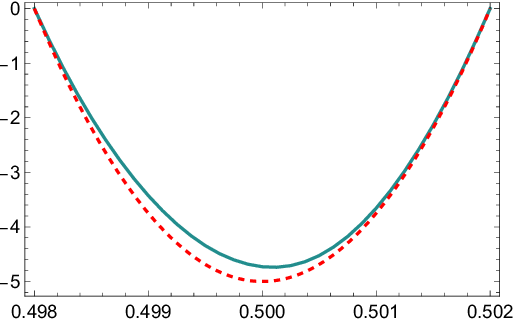}
    \put(48,-7){$x$}
    \put(-6,30){$v$}
    \end{overpic} 
    &   
    \begin{overpic}[width=0.3\textwidth,tics=10]
    {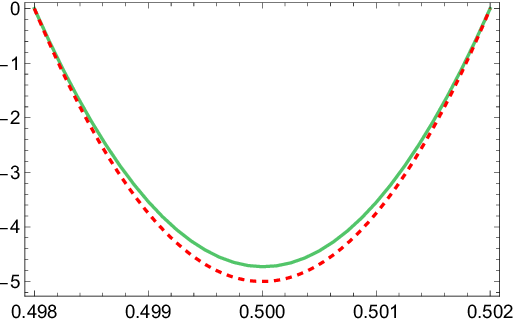}
    \put(48,-7){$x$}
    \end{overpic}
\end{tabular}
\vspace{3mm}
\caption{Numerical solution (in solid lines) and asymptotic parabolic prediction (in red dashed lines) for the $y$-component of velocity, $v$, at the entrance to the centremost branched channel, $x_i = 0.5$, from figure~\ref{fig:FullModel_Flow}. We plot various distances into this region to identify the transition length of the numerical solution into fully-developed flow. Here, $\mathcal{P}_\text{out} = 0.4$, $\Rey = 1000$, $\epsilon = 0.04$, $\delta = 0.1$, $\lambda = 0.1$ and $\gamma = 0.5$.}
\label{fig:TransitionRegion}
\end{figure}

As mentioned in \S\ref{section:BranchedFlow} and \S\ref{section:AngledChannels}, there is a small region near the top of each branched channel where the flow develops into unidirectional, full-developed, flow. We show how the fully developed flow establishes in figure \ref{fig:TransitionRegion}, where we present the velocity, $v$, down the channel with centre located at $x=0.5$, as a function of $x$, at various $y$ positions. We see that the solution has become fully-developed by $y \approx -0.006$, which is $\mathcal{O}(\delta \epsilon)$ for the parameters chosen; the asymptotic prediction for the flow is also shown. We thus conclude that deviations from the asymptotic predictions are confined to an $\mathcal{O}(\delta \epsilon)$ region near the entrance to the \delete{the} branched channel, and may thus be neglected.

\section{}\label{appB}

\begin{figure}
\centering
\begin{overpic}[width=1\textwidth]{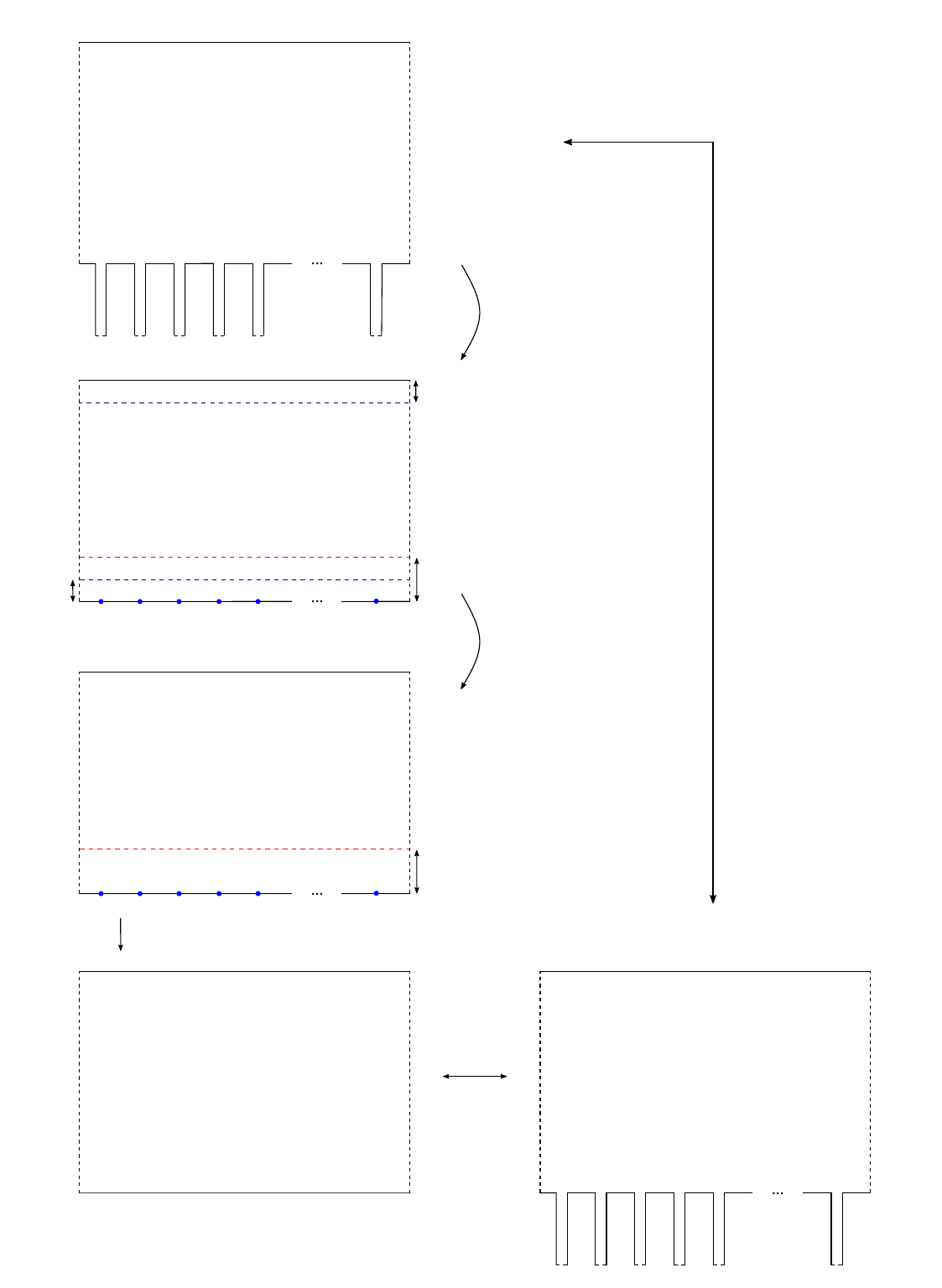}
    \put(14.25,89){Finite $\Rey \gg 1$}
    \put(14.75,86.75){everywhere}
    \put(19.75,95){No-slip on walls}
    \put(14.25,62){Finite $\Rey \gg 1$}
    \put(19.75,69.1){No-slip on walls}
    \put(12,39){Inviscid everywhere}
    \put(22.25,46){Slip on walls}
    \put(12,16){Inviscid everywhere}
    \put(22.25,22.75){Slip on walls}
    \put(50,16){Finite $\Rey \gg 1$}
    \put(50.5,13.75){everywhere}
    \put(58,22.75){Slip on walls}
    \put(57,5.5){\change{Poiseulle}{Poiseuille}}
    \put(59,3.5){flow}
    \put(57.95,1.5){No-slip}
    \put(40,79){Solve flow in}
    \put(38.25,77){branched channels}
    \put(43.75,75){\&}
    \put(41,73){point-sink}
    \put(39.5,71){approximation}
    \put(6.75,31.8){Point-sink strength retains initial $\Rey$}
    \put(39.75,51){Inviscid}
    \put(38.75,49){assumption}
    \put(33,54.5){$\epsilon$}
    \put(33,31.8){$\epsilon$}
    \put(1.5,53.5){$\mathlarger{\frac{1}{\sqrt{\Rey}}}$}
    \put(32.75,69){$\mathlarger{\frac{1}{\sqrt{\Rey}}}$}
    \put(10.5,27){Effective boundary condition, $\epsilon \rightarrow 0$}
    \put(15.25,8.5){$v = -\kappa \mathcal{P}_\text{out}$}
    \put(35.75,14.25){($\star$)}
    \put(56.75,65){($\dagger$)}
    \put(2.75,95.5){(a)}
    \put(2.75,69){(b)}
    \put(2.75,46.5){(c)}
    \put(2.75,23.25){(d)}
    \put(38.25,23.25){(e)}
    \end{overpic}
    \caption{Flow solution structure. The full high-Reynolds-number structure is given by (a), applying several assumptions to find an effective boundary condition (d). We verify the asymptotics of (d) with the appropriate numerical model in (e), denoted by ($\star$). This relates to the original structure (a), via ($\dagger$), retaining the initial $1 \ll \Rey < \infty$ everywhere, whilst imposing an outer inviscid flow assumption via a wall slip condition.}
    \label{fig:Assumption_roadmap}
\end{figure}

In this appendix, we outline the solution procedure taken within this paper and highlight the comparisons we make between numerical simulations and asymptotic predictions, in the ideal and actual comparisons (see figure~\ref{fig:Assumption_roadmap}). 

Our model derivation begins with high-Reynolds-number flow in a branched domain (see figure~\ref{fig:Assumption_roadmap}(a)). We explicitly solve for the flow in each branched channel and then simplify the structure by approximating each branched channel by a point-sink (see figure~\ref{fig:Assumption_roadmap}(b)) --- the strength of each point-sink depends on the finite $\Rey$. We now suppose that the flow in the simplified main channel is inviscid everywhere, which allows us, via a multiple\ownadd{-}scales analysis, to find an effective boundary condition on the bottom wall (see figure~\ref{fig:Assumption_roadmap}(c), (d)) and composite solutions for the velocities.

Since the viscous boundary layers are tiny in this limit, we emulate the flow in the main channel by imposing a slip condition on the walls as shown in figure~\ref{fig:Assumption_roadmap}(e), transforming the problem as indicated by ($\dagger$). The comparison between our asymptotic and numerical solution is 
indicated by ($\star$) in figure~\ref{fig:Assumption_roadmap} enabling us to focus on the influence on the flow of the branched channels.

\section{}\label{appC}

\begin{figure}
\centering
    \begin{overpic}[width=0.6\textwidth,tics=10]{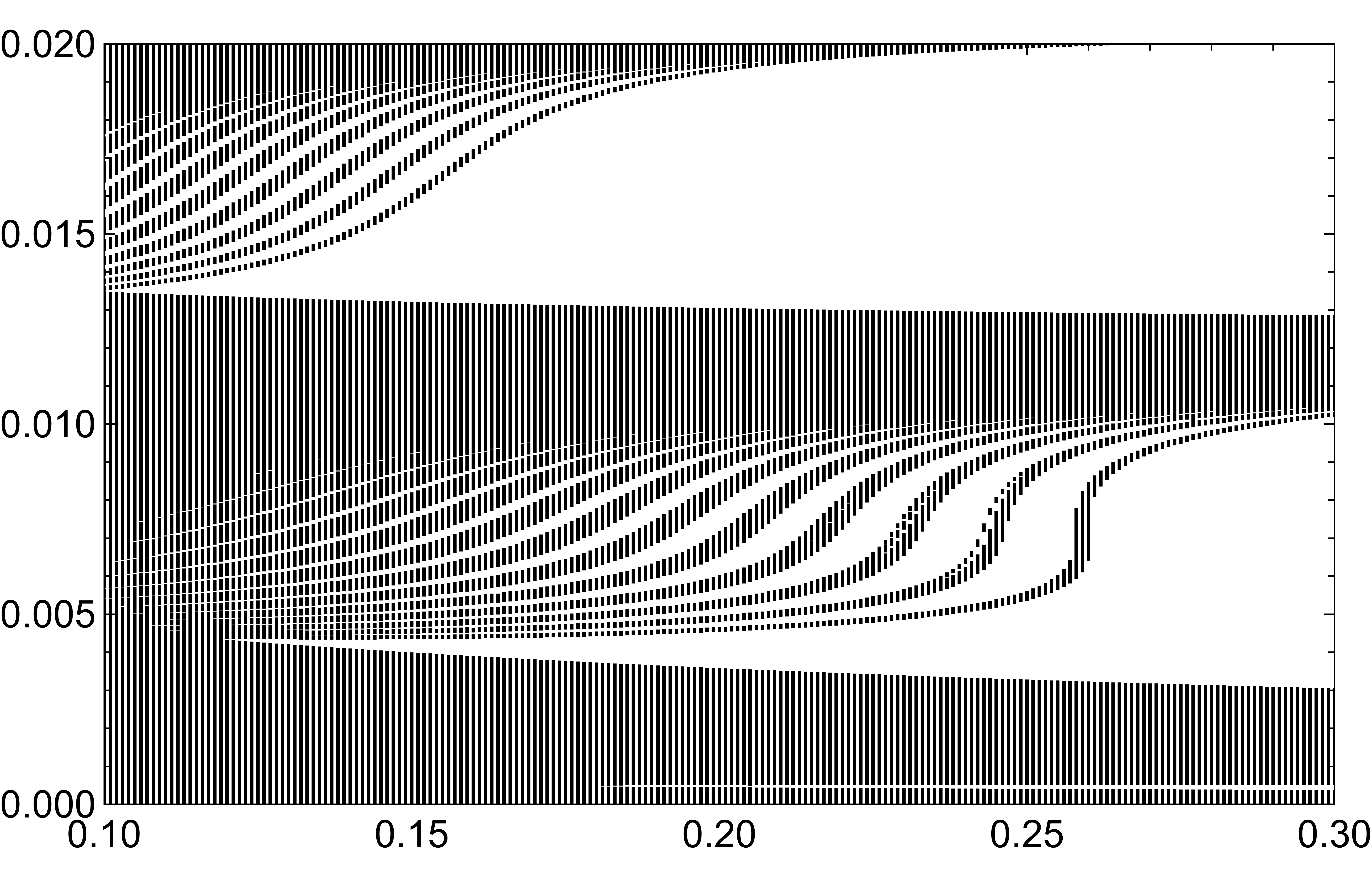}
    \put(50.5,-3.75){$\Sty$}
    \put(-8,33){$y_0$}
    \end{overpic}
\vspace{1mm}
\caption{The inlet positions $y_0$ of particles that exit the device through a branched channel  versus the Stokes number, $\Sty$. For each $\Sty$, we release $i=19,999$ particles at slightly perturbed initial points around $i$ equispaced points, and run $60$ separate simulations, plotting each point once. Here, $\mathcal{P}_\text{out} = 0.4$, $\Rey = 1000$, $\epsilon = 0.04$, $\delta = 0.1$, $\lambda = 0.1$ and $\gamma = 0.5$.}
\label{fig:Grazing}
\end{figure}

Here, we explore in more detail the trajectories that are used to make figures~\ref{fig:KPlots} and \ref{fig:RPlots}. We focus on particle-release positions within the $\epsilon$-boundary layer and we show in figure~\ref{fig:Grazing}, as black points, the $y_0$ locations in each simulation that result in particles going down a branched channel, for a given $\Sty$. We see that, as we increase $\Sty$, tongue-like regions of parameter space open up in which particles do not flow down branched channels. This is most striking for $y_0 < 0.01$. We associate this behaviour with particles hitting the bottom wall very close to the entrance to a branched channel, and we call this process \emph{grazing}. The amount of grazing appears to increase as $\Sty$ increases up to the final point near $\Sty = 0.26$.
\begin{figure}
\centering
\vspace{0mm}
\begin{tabular}{cc}
(a) & (b)\\[4mm]
    \begin{overpic}[width=0.4\textwidth,tics=10]
    {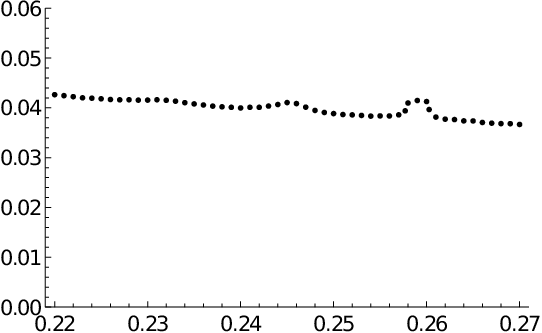}
    \put(50.25,-7){$\Sty$}
    \put(5,67){$\mathcal{K}$}
    \end{overpic}
    &
    \begin{overpic}[width=0.4\textwidth,tics=10]
    {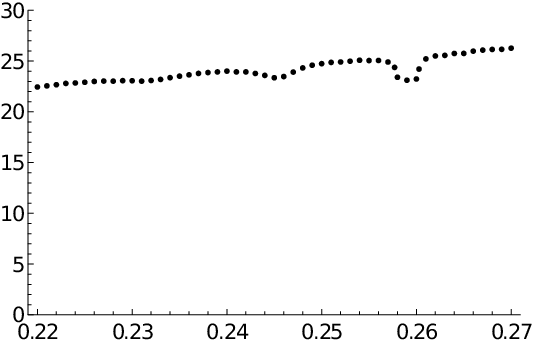}
    \put(50.25,-7){$\Sty$}
    \put(4,67){$\mathcal{R}$}
    \end{overpic}
\end{tabular}
\vspace{2mm}
\caption{Zoomed in versions of figures~\ref{fig:KPlots} and \ref{fig:RPlots} around $\Sty \in [0.22,0.27]$.}
\label{fig:ZoomedKRPlots}
\end{figure}

Comparing figures~\ref{fig:KPlots} and \ref{fig:RPlots} with figure \ref{fig:Grazing}, we notice that the bumps in $\mathcal{K}$ align with the values of $\Sty$ where there is a sharp switching or bifurcation-like behaviour in the $y_0$ values that result in particles passing through the branched channels. We see that the large deviation in $\mathcal{K}$ and $\mathcal{R}$ at $\Sty = 0.26$ is aligned with the steepest change in the $y_0$ values in figure \ref{fig:Grazing}. We note that there are a finite number \add{of} rapidly changing $y_0$ regions. In figure~\ref{fig:ZoomedKRPlots}, we zoom in on the behaviour of $\mathcal{K}$ and $\mathcal{R}$ for $\Sty$ near $0.26$. We see that, with this increased resolution, there are further non-monotonic regions, and that these also correspond to the regions of rapid $y_0$ change in figure~\ref{fig:Grazing}, with the slope here appearing to correlate with the magnitude of the non-monotonic deviations. We hypothesise that the phase-space behaviour discussed in this appendix will depend critically on the precise geometry at the branched channel entrance.

\bibliographystyle{jfm}
\bibliography{jfm}

\end{document}